\PassOptionsToPackage{font=small,bf}{caption}

\documentclass{sig-alternate-05-2015}

\usepackage{graphicx}
\usepackage{url}
\usepackage{amsmath}
\usepackage{amsmath,amsfonts,amssymb,bm,latexsym}
\usepackage{color}
\usepackage{enumitem}
\usepackage{multirow}
\usepackage{colortbl}
\usepackage{subcaption}
\usepackage[font={bf}]{caption}
\definecolor{kugray5}{RGB}{224,224,224}
\usepackage[table]{xcolor}
\usepackage{arydshln}
\usepackage{ctable}
\usepackage{array}
\usepackage{balance}
\usepackage{url}
\usepackage{amssymb}
\usepackage[lined,boxed]{algorithm2e}

\newcommand{\eat}[1]{}
\newcommand{\etal}{\emph{et al.}}
\newcommand{\eg}{\emph{e.g.}, }
\newcommand{\ie}{\emph{i.e.}, }

\newtheorem{theorem}{Theorem}
\newtheorem{definition}[theorem]{Definition}

\newcommand{\thline}{\specialrule{.1em}{.1em}{.1em}}
\newcolumntype{L}[1]{>{\raggedright\let\newline\\\arraybackslash\hspace{0pt}}m{#1}}
\newcolumntype{C}[1]{>{\centering\let\newline\\\arraybackslash\hspace{0pt}}m{#1}}
\newcolumntype{R}[1]{>{\raggedleft\let\newline\\\arraybackslash\hspace{0pt}}m{#1}}

\newcommand{\myparatight}[1]{\smallskip\noindent{\bf {#1}:}~}

\begin{document}
\isbn{n/a}
\doi{n/a}

\title{Structural Analysis of User Choices for Mobile App Recommendation}

\author{
Bin Liu$^1$, Yao Wu$^2$, Neil Zhenqiang Gong$^3$, Junjie Wu$^4$, Hui Xiong$^1$\thanks{Contact Author.}, Martin Ester$^2$ \\
       \affaddr{$^1$Rutgers University, \{binben.liu, hxiong\}@rutgers.edu}\\
       \affaddr{$^2$Simon Fraser University, \{wuyaow,ester\}@sfu.ca}\\
       \affaddr{$^3$Iowa State University, {neilgong@iastate.edu}}\\
       \affaddr{$^4$Beihang University, {wujj@buaa.edu.cn}}\\
}

\maketitle
 
\begin{abstract}

\noindent Advances in smartphone technology have promoted the rapid development of mobile apps. However, the availability of a huge number of mobile apps in application stores has imposed the challenge of finding the right apps to meet the user needs. Indeed, there is a critical demand for personalized app recommendations. Along this line, there are opportunities and challenges posed by two unique characteristics of mobile apps. First, app markets have organized apps in a
hierarchical taxonomy. Second, apps with similar functionalities are competing with each other. While there are a variety of approaches for mobile app recommendations, these approaches do not have a focus on dealing with these opportunities and challenges. To this end, in this paper, we provide a systematic study for addressing these challenges. Specifically, we develop a \emph{structural user choice model} (SUCM) to learn fine-grained user 
preferences by exploiting the hierarchical taxonomy of apps as well as the competitive relationships among apps. Moreover, we design an efficient learning algorithm to estimate the parameters for the SUCM model. Finally, we perform extensive experiments on a large app adoption data set collected from Google Play. The results show that SUCM consistently outperforms state-of-the-art
top-N recommendation methods by a significant margin.

\end{abstract}


\section{Introduction}

Recent years have witnessed the tremendous growth in mobile devices among an increasing number of users, and the penetration of mobile devices into every component of modern life. 
Indeed, the smartphone market surpassed the PC market in 2011 for the first time in history\footnote{{The Smartphone Market is Bigger Than the PC Market} (2011),~ \url{http://www.businessinsider.com/smartphone-bigger-than-pc-market-2011-2}}.
Thereafter, the smartphone market has continued to increase dramatically, {\it e.g.}, the smartphones shipped in the third quarter of 2013 increased 44\% year-on-year\footnote{{Smartphone Sales in the Third Quarter of 2013} (2013),~\url{http://www.finfacts.ie/irishfinancenews/article_1026800.shtml}}. 
One of the reasons lies in the fact that users are able to augment the functions of mobile devices by taking advantage of various feature-rich third-party \emph{applications} (or apps for brevity), which can be easily obtained from centralized markets such as Google Play and App Store. However, the availability of a huge number of mobile apps in application stores has imposed the challenge of finding the right apps to meet the user needs. For instance,  as of July 2013, Google Play had over 1
million apps with over 50 billion cumulative downloads, and the number of apps had reached over 1.4 million in January 2015\footnote{{Google Play Statistics}, Retrieved January 2015,~\url{http://en.wikipedia.org/wiki/Google_Play}}; as of February  2015, App Store had over 1.4 million apps and a cumulative of over 100 billion apps downloaded\footnote{{App Store Statistics}, Retrieved January 2015,~\url{http://en.wikipedia.org/wiki/App_Store_(iOS)}}. As a result, there is a critical demand for effective personalized app recommendations.

However, for the development of personalized app recommender systems, there are opportunities and challenges posed by two unique characteristics of mobile apps. First, application stores have organized apps in a hierarchical taxonomy. For instance, Google Play groups the apps into 27 categories, such as  {\it social}, {\it games}, and {\it sports} according to their functionalities. These categories can be further divided into subcategories, \eg~apps in the category of {\it games} are further divided into subcategories such as {\it action}, {\it arcade}, and {\it puzzle}. For the apps in the same category or subcategory, they have similar functionalities. Then, how a user navigates through the hierarchy to locate relevant apps represents a fine-grained interest preference of the user. Thus, the first challenge is how to leverage this hierarchical taxonomy of apps to better profile user interests and enhance app recommendations. Second, apps with similar functionalities are competing with each other. For instance, when a user has already adopted Google Maps as his/her navigation tool, the user might not be interested in other navigation tools such as Apple Maps. While there are a variety of existing approaches for mobile app recommendations, these approaches do not have a focus on dealing with these opportunities and challenges.

Instead, in this paper, we provide a systematic study to address these challenges. Specifically, 
we first develop a \emph{structural user choice model} (SUCM) to learn  fine-grained user preferences by exploiting the hierarchical taxonomy of apps as well as the competitive relationships among apps\footnote{Note that the model and algorithms developed in this paper can also be applied to other domains where items are organized in a hierarchical way.}. Since apps are organized as a  hierarchical taxonomy, we model the user choice as two phases. In the first phase,
a user decides which type of apps to choose and then moves to the appropriate app category/subcategory. In the second phase, the user chooses apps in the selected category/subcategory. Such structural user choice is modeled by a unique \emph{choice path} over the tree hierarchy, where the choice path starts from the root of the hierarchy and goes down to the app that is selected by a user. In each step of moving along the choice path, the competitions between the
candidates (\emph{i.e.}, either the same level categories/subcategories or apps in a chosen category/subcategory) play an important role in affecting user's choices. We capture the structural choice procedure by cascading user preferences over the choice paths through a probabilistic model. 
Specifically, in our probabilistic model, motivated by the widely used discrete choice models in economics~\cite{Luce:ChoiceAxiom1959,McFadden:choide:1973,Takeshi:ChoiceModel:1977}, we model the probability that a user reaches a certain node in the choice path as a \emph{softmax} of the user's preference on the chosen node over the user's preference on all the nodes at the sample level.
The softmax function is used to capture the competitions between categories/subcategories or  apps in a category/subcategory. Moreover, we model a user's preference over one node using latent factors, which enables us to capture the correlations between nodes. 

Moreover, we design an efficient learning algorithm to estimate the parameters of the SUCM model. 
The major challenge of learning the parameters lies in the softmax on the leaf nodes (apps) of the tree hierarchy. Indeed, it is impractical to optimize these softmax functions for a subcategory of apps by directly applying Stochastic Gradient Descent (SGD), because the time complexity of one SGD step is linear to the number of apps under the subcategory, which might be very large. 
To address this challenge, we relax the softmax term in each subcategory into a hierarchical softmax, thus the time complexity of learning parameters is reduced to be logarithm of the number of apps under the subcategory.

Finally, we collected a large-scale dataset from Google Play to evaluate our approach and compare SUCM with state-of-the-art approaches. The experimental results show that SUCM consistently outperforms these methods with a significant margin in terms of a variety of widely used evaluation metrics for top-N recommendation.

\def \uu {{\bf u}}
\def \vv {{\bf v}}
\def \pp {{\bf p}}
\def \qq {{\bf q}}
\def \YY {{\bf Y}}
\def \UU {{\bf U}}
\def \VV {{\bf V}}
\def \PP {{\bf P}}
\def \QQ {{\bf Q}}
\def \RR {\mathbb{R}}
\def \Pr {{\mathrm{Pr}}}
\def \DD {{\mathcal D}}
\def \Poi {{\mathrm{Poisson}}}
\def \II {{\bf I}}

\begin{figure*}[!t]
\centering
\includegraphics[width=0.8\textwidth]{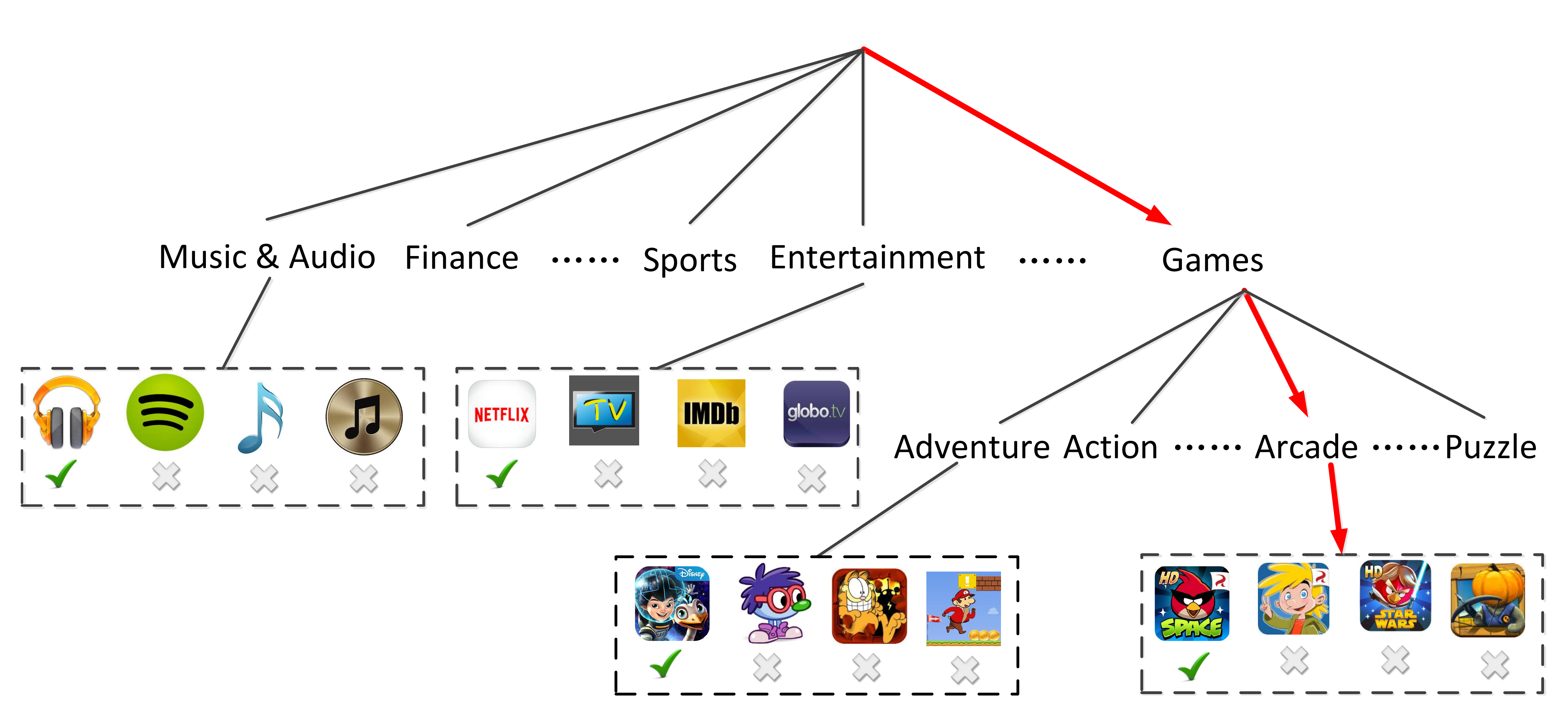}
\caption{An illustrative example of structural user choice for app adoption in Google Play. First, apps are organized into a Category Tree. Second, as illustrated by the highlighted and arrowed path, a user makes an app adoption by traversing a \emph{choice path} from the root of the tree to the chosen app.}
\label{Nested_tree}
\end{figure*}

\section{Problem Definition}
We first introduce three key concepts and then formally define our app recommendation problem.

\begin{definition}[Category Tree]
    A Category Tree (denoted as $\Gamma$) is a data structure to organize apps according to their properties (\eg~functionalities).  Figure \ref{Nested_tree} shows an example Category Tree adopted by Google Play. In a Category Tree,  internal nodes represent categories or subcategories,  leaf nodes represent apps, and the children of an internal node represent the subcategories or the apps that belong to the category/subcategory represented by the node. We use $z$ to denote an internal node in $\Gamma$, and we denote by $level(z)$, $c(z)$, $\pi(z)$, and $s(z)$ the level, the children, the parents, and  the siblings of $z$, respectively. Moreover, we use $z_M$ to denote an internal node whose children are leaf nodes and use $i$ to represent an app.
\end{definition}

We note that an app might belong to multiple categories due to the rich functionalities provided by it, which makes the category hierarchy not a tree. However, we found that mobile markets such as Google Play do not place an app into multiple categories based on the dataset we collected from Google Play, and thus we do not consider this scenario.

\begin{definition}[Choice Path]
    A choice path is a sequence of nodes that a user traverses through the Category Tree $\Gamma$, starting from the root and ending at a leaf node which corresponds to the app selected by the user. For instance, if a user adopts an app $i$, the choice path can be represented as $\mathrm{path}_{i} = z_0 \rightarrow z_1 \rightarrow \cdots \rightarrow z_M \rightarrow i$. Note that, given the Category Tree $\Gamma$, the choice path $\mathrm{path}_{i}$ for app $i$ is unique.
\end{definition}

\begin{definition}[Competing Apps]
For an app $i$, we denote by $\mathcal A(i)$ the set of apps that have competing properties (\eg~functionalities) and compete with $i$ to attract users. In this paper, we treat the siblings of an app $i$ under the same category/subcategory in the Category Tree as the competing apps. 
\end{definition}

We note that users might have multiple ways to adopt apps, \eg 
suggestions from friends, recommendations from Google Play store, etc. 
However, we assume that no matter in which way a user is aware of an app, 
the decision is made on the functionality of the app and its competitors with similar functionalities, 
thus following the choice path we discuss above.   

It also should be noted that we do not assume a user only adopts one app in a subcategory. 
The category/subcategory in the Category Tree provided by mobile markets such as Google Play is not fine-grained enough so some siblings of the app $i$ might provide slightly different functionalities with $i$. 
For example, Facebook, LinkedIn and Twitter all belong to {\it Social} category. 
We model the process of one user adopting an app using a structural choice model. If a user selects multiple apps under a same category, the joint probability of selecting them together would be optimized (see details in Section \ref{sec:model}).

\vspace{5pt}
Given the above three concepts, we can formally define our app recommendation problem as follows: suppose we are given a set of users denoted as $\mathcal U =\{1, 2, ..., U\}$, a set of apps denoted as  $ \mathcal I =\{1, 2, ..., I\}$, the apps are organized into a predefined  Category Tree $\Gamma$, each app $i$ has a set of competing apps $\mathcal A(i)$, a set of adoption records $\{(u, i)\}$  indicating which users have adopted which apps, then our goal is to recommend each user a list of apps that match his/her interest preference. 
In the rest of the paper, we use $u$ to index users, and $i$ and $j$ to index apps. Moreover, we use the two terms \emph{app} and \emph{item} interchangeably. Table \ref{math_notation} shows some important notations used in this paper.

\begin{table*}[th]
\caption{Mathematical Notations}
\label{math_notation}
\vspace{-10pt}
\begin{center}
 \begin{tabular}{ C{3cm}  L{8cm}} \thline
    \textbf{Symbol}  & \textbf{Description} \\ \thline
    $u$ &  user index for user set $\mathcal U =\{1, 2, ..., U\}$ \\ \hline
    $i, j$ &  app index for app set $ \mathcal I =\{1, 2, ..., I\}$ \\  \hline
    $\Gamma$ &   predefined Category Tree \\ \hline
    $z$ &  internal node in Category Tree $\Gamma$, in particular, $z_M$ denotes a node whose children are leaf nodes \\ \hline
    $\mathrm{path}_{i} $ & choice path in $\Gamma$:  $z_0 \rightarrow z_1 \rightarrow \cdots \rightarrow z_M \rightarrow i$ \\ \hline
    $\pi(z), s(z), c(z)$ & parent, sibling, and children of internal node $z$ in the Category Tree $\Gamma$\\  \hline
    $\pp_u, \qq_i, \qq_z$ & latent factor vector for user $u$, app $i$, and internal node $z$ in  $\Gamma$\\  \hline
    $b_i, b_z$ & bias term for app $i$ and  internal node $z$\\  \hline
    $y_{ui}$ & affinity score of user $u$ for app $i$\\  \hline
    $y_{uz}$ & affinity score of user $u$ for internal node $z$\\ \hline
    $\mathcal D = \{(u,i)\}$ & observed user-app adoption instances \\ \hline
    $\mathcal{D}_u$ & adopted apps by user $u$ \\ \thline
    \end{tabular}
\end{center}
\end{table*}

\section{Structural User Choice Model} 
\label{sec:model}
In this section, we present our \emph{structural user choice model} (SUCM) to learn fine-grained user interest preference via leveraging the Category Tree and  competitions between apps for app recommendation.

\subsection{Model Structural User Choice}
As shown in Figure \ref{Nested_tree}, given a Category Tree  $\Gamma$, there exists one unique {\it choice path} from the root node to app $i$, namely, 
$$\mathrm{path}_{i} = \underbrace{z_0 \rightarrow z_1 \rightarrow \cdots \rightarrow z_M}_{\substack{\text {Phase I:}\\\text{locate a subcategory}}} \underbrace{\longrightarrow  i.}_{\substack{\text {Phase II:}\\\text{choose an app}}}$$
We see the structural user choice consists of two adoption phases. 
In the first phase, a user decides what types of apps to choose and moves to the appropriate category or subcategory in the Category Tree, namely, traverses $z_0 \rightarrow z_1 \rightarrow \cdots \rightarrow z_M$.  In the second phase, the user makes app adoption decisions by choosing app $i$ among all competing apps under the located subcategory $z_M$. 
For example, if a user wants to select the app {\it Angry Birds} under the subcategory  {\it Arcade}, he would first consider the {\it Games} category and then further locates himself at the {\it Arcade} subcategory before he finally chooses  app {\it Angry Birds}. 

We model the process of a user $u$ traversing path $z_0 \rightarrow z_1 \rightarrow \cdots \rightarrow z_M \rightarrow i$ as a sequence of decisions made for the multiple competing choices at each choice step. Specifically, in each step among this decision-making sequence:
\begin{itemize} 
\setlength{\itemsep}{5pt}
\item for choosing category or subcategory, user $u$ chooses one child node $z$ from all the children $c(\pi(z))$ of $z$'s parent node $\pi(z)$; 
\item for choosing app, user $u$ chooses app $i$ from all the children of $i$'s parent node, namely $z_M$. 
\end{itemize}
Each decision making step can be seen as a discrete choice model, whose theoretical foundation is the neoclassical economic
theory on preferences and utility built on a set of axiomatic assumptions~\cite{Luce:ChoiceAxiom1959,McFadden:choide:1973,Takeshi:ChoiceModel:1977}. 
The discrete choice model implies that a user $u$ is endowed a {\it utility value} $f(u,z)$ to each alternative $z$ in a choice set $\mathcal A(z)$. In our recommendation task, the utility value $f(u,z)$ can be the affinity score, which captures user preferences, between user $u$ and choice $z$. Following the random utility model~\cite{Takeshi:ChoiceModel:1977},  we model the utility as a random variable
\begin{equation}
\nu_{uz}  = f(u,z) + \varepsilon_{uz},
\end{equation}
where $f(u,z)$ is the deterministic part of the utility reflecting user preference, and $\varepsilon_{uz}$ is the stochastic part capturing the impact of all unobserved factors that affect the user's choice.  By assuming the stochastic part $\varepsilon_{uz}$ be an independently and identically distributed log Weibull (type I extreme value)  distribution, we can obtain the the multinomial choice model \cite{McFadden:choide:1973}. Specifically, in a multinomial choice model, the probability of a user  $u$ choosing $z$ from a choice set $\mathcal A(z)$takes the form of 
\begin{equation}\label{equ:choice_model}
\begin{split}
\Pr(\mathrm{user~} u \mathrm{~choose~} z|\mathcal A(z)) = \frac{\exp(f(u,z))}{\sum_{z'\in \mathcal A(z)}\exp(f(u,z'))},
\end{split}
\end{equation}
where $f(u,z)$ is a user preference depended utility function. This choice model also holds for user $u$ choosing app $i$ from app choice set $\mathcal A(i)$. Note that the choice model $\frac{\exp(f(u,z))}{\sum_{z'\in \mathcal A(z)}\exp(f(u,z'))}$ turns out to be a {\it softmax} function of utility value $f(u,z)$. 
In the following, we elaborate how we model each phase.  

\vspace{10pt}
\noindent{\bf Phase I: Model category/subcategory preference.}
Following the latent factor models that are widely used in  conventional recommender systems~\cite{PMF:NIPS08,Koren:FMN:KDD08}, we use a latent factor vector $\pp_u \in \mathbb R^K$ to represent a user's latent interest, where $K$ is the dimension of the latent factor vector. 
Intuitively, $\pp_u $ captures the interest of the user $u$.
To capture the hierarchical structural user choice, we associate an internal node $z$ in the Category Tree with a latent factor vector $\qq_z$, which represents the properties (\eg~functionalities) of $z$ in the latent space. Moreover, we define the affinity score between a user $u$ and an internal node $z$ as
\begin{equation}
y_{uz} = b_z +  \pp_u^\top  \qq_z,
\end{equation}
where $b_z$ is a bias term for the node $z$. 
The category/subcategory node affinity score represents the preference of a user over the category or the subcategory of apps (\eg~Games).

We model the process of a user locating a subcategory as a sequence of decisions made for the multiple competing choices, starting from the root node and moving along the Category Tree towards the internal node corresponding to the subcategory. 
Specifically, in each step among this decision-making sequence, user $u$ chooses one child node $z$ from all the children of $z$'s parent node $\pi(z)$. Following the choice model as shown in Equation (\ref{equ:choice_model}), we assume the utility as the affinity score between user $u$ and  internal node $z$, \ie $$f(u,z)=y_{uz} = b_z +  \pp_u^\top  \qq_z.$$
Then we model the probability of user $u$ choosing the child $z$ from all the children $c(\pi(z))$ of $z$'s parent node $\pi(z)$ as a \emph{softmax} function of the affinity scores between the user $u$ and the internal nodes $c(\pi(z))$. Formally, we have:
\begin{equation}
\begin{split}
\Pr(z|u, \pi(z)) &= \frac{\exp(y_{uz})}{\sum_{z'\in c(\pi(z))}\exp(y_{uz'})} \\
\end{split}
\end{equation}
The \emph{softmax} function is used to model the competitions between the nodes in $c(\pi(z))$. As a result, the probability of user $u$ traverses $z_0 \rightarrow z_1 \rightarrow \cdots \rightarrow z_M$ to reach the subcategory $z_M$ is cascaded as
\begin{equation}
\begin{split}
 \Pr(z_0 \rightarrow z_1 \rightarrow \cdots \rightarrow z_M|u) 
= \prod_{m=1}^{M} \Pr(z_m|u, z_{m-1}) 
\end{split}
\end{equation}

\vspace{10pt}
\noindent{\bf Phase II: Model app adoption.} 
After a user locates at a specific subcategory node $z_M$ whose children are all apps, the user makes an app adoption decision by choosing an app $i$ among all competing choices $c(z_M)$. 
We use a latent factor vector $\qq_i \in \mathbb R^K$ to represent the latent factor of app $i$.  
Intuitively, $\qq_i$ encodes the properties (\eg~functionalities) of app $i$.  Moreover, we define the affinity score between user $u$ and app $i$ as
\begin{equation}
y_{ui} = b_i + \pp_u^\top \qq_i,
\end{equation}
where $b_i$ is a bias term for app $i$. Again, following the choice model as shown in Equation (\ref{equ:choice_model}), we assume the utility as the affinity score between user $u$ and  app $i$, \ie $$f(u,i)=y_{ui} = b_i +  \pp_u^\top  \qq_i.$$
Then we model the probability of user $u$ selecting app $i$ over its competing alternatives under the subcategory node $z_M$ using a softmax function as follows: 
\begin{equation}
\begin{aligned}
    \Pr(i|u, z_M) & = \frac{\exp(b_i + \pp_u^\top \qq_i )}{\sum_{j\in c(z_M)}\exp(b_j + \pp_u^\top \qq_j)},
\end{aligned}
\end{equation}
where $z_M$ is the parent node of app $i$ and $c(z_M)$ includes all competing apps of app $i$ and $i$ itself. The softmax function is used to model the competitions between apps. 

\vspace{10pt}
\noindent{\bf Model the overall structural choice probability.}
Note that there exists one unique {\it choice path} from the root node to app $i$, namely, $$\mathrm{path}_{i} = z_0 \rightarrow z_1 \rightarrow \cdots \rightarrow z_M \rightarrow i.$$
Then, the probability of user $u$ choosing app $i$ is the joint probability of $u$ selecting each node in the choice path $\mathrm{path}_{i}$, \emph{i.e.}, we have:
\begin{equation}
\begin{split}
 \Pr(i|u) &= \Pr(i|u, z_M) \Pr(z_0 \rightarrow z_1 \rightarrow \cdots \rightarrow z_M|u) \\
 &= \Pr(i|u, z_M) \prod_{m=1}^{M} \Pr(z_m|u, z_{m-1}) 
\end{split}
\end{equation}
where the first term $\Pr(i|u, z_M)$ is user $u$'s adoption probability of app $i$ under subcategory node $z_M$ and the second term $\prod_{m=1}^{M} \Pr(z_m|u, z_{m-1})$ captures the structural choice by cascading user preferences over the Category Tree $\Gamma$.

\subsection{Model Structural App Dependences}
Intuitively, nodes that are closer in the Category Tree $\Gamma$ could have more similar  properties. For instance, apps under the subcategory {\it action} are more similar to those under the subcategory {\it arcade} than those under the category  {\it weather} because both {\it action} and  {\it arcade}  belong to  the {\it games} category. Thus, we associate each internal node $z$ with a latent variable $\qq_z$ to represent the category/subcategory level properties, and we model the latent variable $\qq_z$ as a function of the latent variable of $z$'s parent node $\qq_{\pi(z)}$ to capture the hierarchical structural  dependences between the nodes in the Category Tree. Formally, we have:
\begin{equation}
    \qq_z \sim \left\{ 
  \begin{array}{l l}
    \mathcal N(0, \sigma^2 \II) & \quad \text{if $z$ is the root node}\\
    \mathcal N(\qq_{\pi(z)}, \sigma^2 \II) & \quad \text{otherwise} 
  \end{array} \right.
\end{equation}
where  $\mathcal N(u, \sigma^2)$ is a normal distribution with mean $u$ and standard deviation $\sigma$. 

\subsection{Discussion} 
Note that our model does not only capture the competitions between apps under the same categories, but also incorporates the correlations between apps via the latent factor representations.
\begin{itemize}
\setlength{\itemsep}{5pt}
\item{\bf Competition}. We use a softmax function to model the probability of selecting a child node (a subcategory or an app) under a category node. If user $u$ selects a child node $z$ from all the competing nodes $\mathcal A(z)$, the value of $y_{uz}$ should be larger than all other $y_{u z^\prime}$ where $z^\prime \in \mathcal A(z)$ and $z \neq z^\prime$. This model characteristic can address the cases when multiple apps in same categories are adopted. 
\item{\bf Correlation}. The latent factor model is able to model the correlations between apps and categories. 
        For example, if two categories are always liked by the same users, the latent factors of them will be close to each other in the latent space.  
        As a result, if we know a user likes one of the two categories, the value of his/her preferences on the other one will also be large. 
\end{itemize}

Here we highlight some important differences between our model and previous work (see Section \ref{sec:baseline} and Section \ref{sec:comp} for details): 
\begin{itemize}
    \item Instead of fitting a point-wise regression model (\eg PMF~\cite{PMF:NIPS08} and LLFM~\cite{Agarwal:KDD2009:RLFM}), the proposed model SUCM optimizes the choice decision making through choice probabilities cascaded in the hierarchy structure. 
    \item Previous feature-based latent factor approaches (\eg SVDFeature~\cite{SVDfeature:2012} and LibFM~\cite{libFM:2012,rendle2010factorization}) utilize item features by representing the user preference on an item using a linear combination of the user-item affinities and the user-feature affinities. Differently, SUCM is designed for structurally organized features, and it models the structure by cascading the choice probabilities instead of linear combinations. 
    \item SUCM generalizes the {\it flat} choice model -- Collaborative Competitive Filtering (\textbf{CCF})~\cite{Yang:sigir2011:CCF} -- to a {\it structural} choice model via leveraging the hierarchy information. We also present an efficient learning algorithm based on Hierarchical Softmax (see Section \ref{sec:hs}) that can also be used for CCF.
\end{itemize}

\section{Parameter Estimation}
Let $\Theta = \{ \pp_u, \qq_i, \qq_z, b_i, b_z \}_{u\in \mathcal U, i\in \mathcal I, z\in \Gamma} $ denote all parameters to be estimated. Given the observed user-app adoption records $\mathcal D = \{(u, i, \mathrm{path}_i)\}$ and the category tree $\Gamma$, we have the posterior probability distribution of the parameters as follows: 
\begin{equation}
\begin{split}
    \Pr(\Theta|\mathcal D, \Gamma)  \propto  & \prod_{u=1}^U \prod_{i \in \mathcal{D}_u} \Pr(i|u, z_M) \prod_{m=1}^{M} \Pr(z_m|u, z_{m-1}) \\ 
    &  \times \prod_{\substack{m=1 \\\forall z\in \Gamma}}^{M} \Pr(\qq_{z_m}|\qq_{z_{m-1}}, \sigma^2 \II) 
\end{split}
\end{equation}
where the first term captures the structural user choices and the second term represents the hierarchical structural dependences of the nodes in the category tree.  We  estimate all the parameters via maximizing the log likelihood of the posterior, 
\begin{equation}\label{equ:log_like}
\begin{split}
    \underset{\Theta}{\operatorname{arg\,max}} & \Bigg\{ \sum_{u=1}^U \sum_{i \in \mathcal{D}_u}  \ln \Pr(i|u, z_M) + \sum_{u=1}^U \sum_{i \in \mathcal{D}_u}\sum_{m=1}^{M} \ln \Pr(z_m|u, z_{m-1}) \\
 & + \sum_{\substack{m=1 \\\forall z\in \Gamma}}^{M} \ln \Pr(\qq_{z_m}|\qq_{z_{m-1}}, \sigma^2 \II)\Bigg\}.
\end{split}
\end{equation}
Note that the widely used regularizations for latent factor vectors~\cite{PMF:NIPS08,Koren:SVDRec2009} can be applied here, but we exclude the regularization priors for presentation simplicity. 

\subsection{Hierarchical Softmax}
\label{sec:hs}

One challenge of directly solving the objective function as shown in Equation (\ref{equ:log_like}) rests in the updating of all the parameters over the probability distribution $\Pr(i|u, z_M)$, namely, the first term in Equation (\ref{equ:log_like}), 
\begin{equation}\label{softmax_fun}
\begin{split}
    & \sum_{u=1}^U \sum_{i \in \mathcal{D}_u} \ln \Pr(i|u, z_M) \\
     = & 
    \sum_{u=1}^U \sum_{i \in \mathcal{D}_u} \ln \frac{\exp(b_i + \pp_u^\top \qq_i)}{\sum_{j\in c(z_M)}\exp(b_j + \pp_u^\top \qq_j)}\\
    = & \sum_{(u, i )\in \mathcal D}\Bigg\{ \left(b_i + \pp_u^\top \qq_i\right) - \ln \Bigg[ \sum_{j \in c(z_M)}  \exp \left(b_j +\pp_u^\top \qq_j\right) \Bigg] \Bigg\}, 
\end{split}
\end{equation}
where $c(z_M)$ represents all the apps under the subcategory $z_M$.
The updating computation cost for all the  parameters in one user-app adoption instance $(u,i)$ is linear to the number of apps under $z_M$, which might be very large.
To address this challenge, we leverage \emph{hierarchical softmax} to approximate $\Pr(i|u, z_M)$ efficiently. 
Hierarchical softmax was first introduced by Morin and Bengio~\cite{hierarchical_softmax:2005} 
for neural networks and recently was widely used in deep learning~\cite{mikolov:NIPS2013:word2vec,mikolov2013efficient}.
The main advantage of hierarchical softmax is that, in each training instance, instead of evaluating the parameters for all the children of $z_M$, 
we only need to evaluate parameters for $\log |c(z_M)|$ nodes.

\begin{figure}
    \centering
    \includegraphics[width=0.5\textwidth]{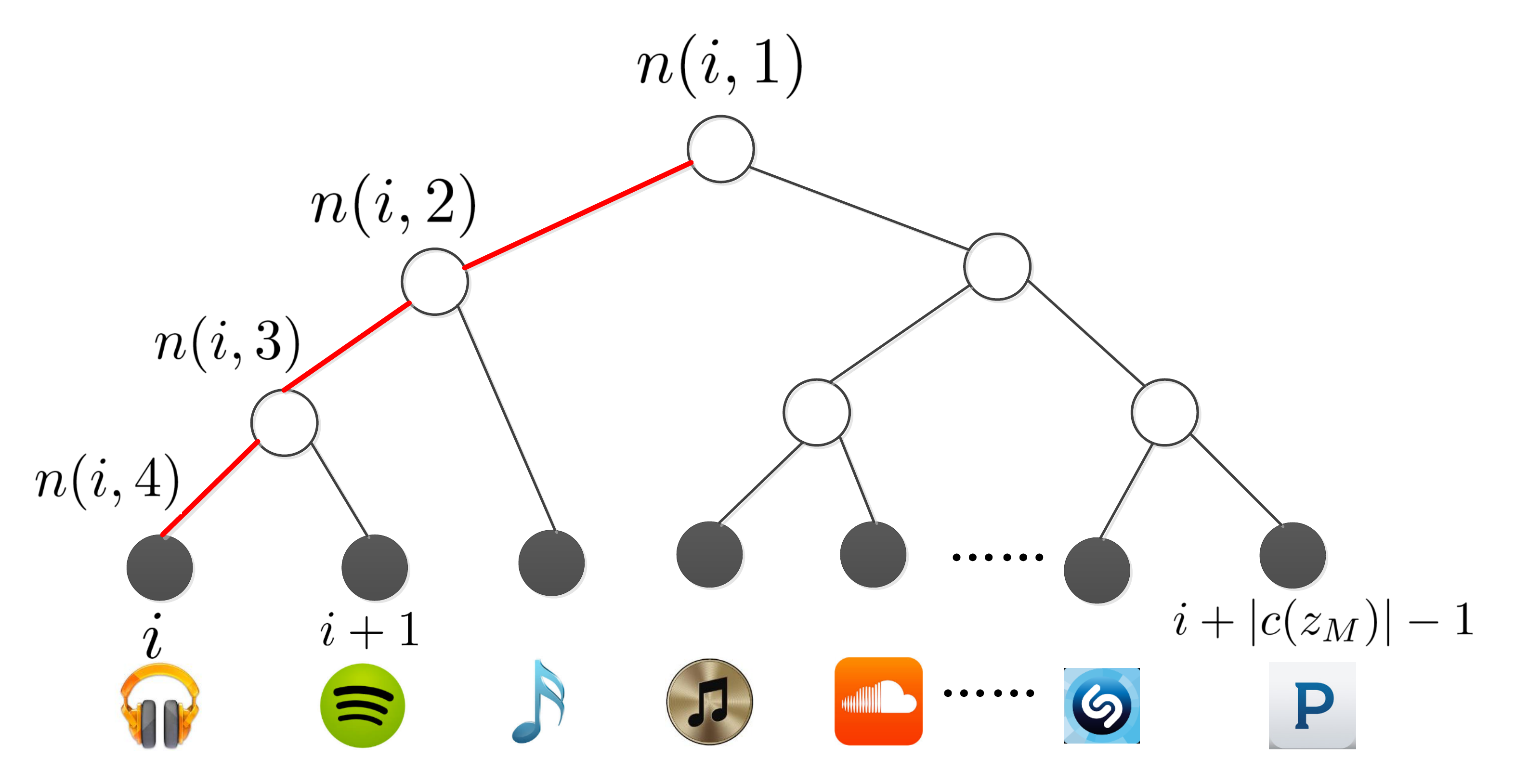}
    \caption{An illustrative example of binary tree for hierarchical softmax under a category/subcategory. All apps under the category/subcategory (\eg Music \& Audio) are organized using a binary tree. The black nodes (leaf nodes) are apps, and the white nodes are internal nodes. One example path from root node to app $i$ is highlighted as $n(i,1) \rightarrow n(i,2) \rightarrow  n(i,3) \rightarrow n(i, 4)$, which means the path length $L(i)=4$. }\label{fig:hs_tree}
\end{figure}

Adapting hierarchical softmax to our model is challenging since our hierachical category tree has multiple layers and applying hierarchical softmax to different 
layers results in different performances. 
In our work, since the major computation cost comes from the large number of apps, we adapt hierarchical softmax to the apps. Specifically, we organize the apps under a  subcategory using a binary tree. 
As shown in Figure \ref{fig:hs_tree}, we represent each app (black nodes) as a leaf node of the binary tree, and the leaf nodes are connected by internal nodes (white nodes).
Let $n(i,l)$ be the $l$-th node on the path from the root of the binary tree to $i$, 
and let $L(i)$ be the length of this path, then $n(i,1) $ is the root and $n(i, L(i)) = i$. 
For each leaf node (\emph{i.e.}, an app), there exists an unique path from the root to the node.  
Let $n(i,l+1) = \mathrm{left}(n(i,l))$ indicate that $n(i,l+1)$ is the left child node of $n(i,l)$ and we define a  sign function as follows:
\begin{equation}
\mathbb{S}\big(n(i,l+1) = \mathrm{left}(n(i,l))\big) :=
\begin{cases} 
1 & n(i,l+1)~ \text{on left}, \\
-1 &\text{otherwise}.
\end{cases}
\end{equation}

Let  $y_{u,n(i,l)}$ be the affinity score between user $u$ and node $n(i,l)$, which is defined as
\begin{equation}
    y_{u,n(i,l)} = b_{n(i,l)} + \pp_u^\top \qq_{n(i,l)},
\end{equation}
where $\qq_{n(i,l)} \in \mathbb{R}^K$ is the latent factor vector and $b_{n(i,l)}$ is the bias term for node $n (i,l)$. 
Intuitively, at each inner node $n(i,l)$ in the hierarchical softmax binary tree, we assign the probability of moving left as 
\begin{equation}\label{equ:left}
\begin{aligned}
\Pr\big(u,n(i,l+1) = \mathrm{left}(n(i,l))\big) = \sigma\big(b_{n(i,l)} + \pp_u^\top \qq_{n(i,l)}\big),
\end{aligned}
\end{equation}
where $\sigma(x)$ is a sigmoid function defined as follows:
\begin{equation}
\sigma(x) = \frac{1}{1 + e^{-x}}. 
\end{equation}
Accordingly, the probability of moving right is 
\begin{equation}\label{equ:right}
\begin{aligned}
    \Pr\big(u,n(i,l+1) \neq \mathrm{left}(n(i,l))\big) & = 1- \sigma\big(b_{n(i,l)} + \pp_u^\top \qq_{n(i,l)}\big) \\
& = \sigma \Big(-\big(b_{n(i,l)} + \pp_u^\top \qq_{n(i,l)}\big)\Big).
\end{aligned}
\end{equation}
Combing Equation (\ref{equ:left}) and Equation (\ref{equ:right}), we can derive from the probability of moving from node $n(i,l)$ to node $n(i,l+1)$ as 
\begin{equation}
\begin{aligned}
    & \Pr\big(u, n(i,l) \rightarrow n(i, l+1)\big) \\
    = & \sigma \Big(\mathbb{S}\big(n(i,l+1) = \mathrm{left}(n(i,l))\big)\cdot \big(b_{n(i,l)} + \pp_u^\top \qq_{n(i,l)}\big) \Big ).
\end{aligned}
\end{equation}

As a result, by using the path $n(i,1) \rightarrow n(i,2) \cdots \rightarrow n(i, L(i))$ in the defined hierarchical softmax binary tree, we  approximate the probability $\Pr(i|u, z_M)$ as follows:
\begin{equation}\label{h_softmax_fun}
    \begin{aligned}
         & \Pr(i|u, z_M)  =   \prod_{l=1}^{L(i)-1} \Pr({u,n(i,l)\rightarrow n(i,l+1)}) \\
         = & \prod_{l=1}^{L(i)-1} \sigma \Big(\mathbb{S}\big(n(i,l+1) = \mathrm{left}(n(i,l))\big)\cdot \big(b_{n(i,l)} + \pp_u^\top \qq_{n(i,l)}\big) \Big ).
\end{aligned}
\end{equation}

Note that, instead of computing the affinity scores for all the apps under subcategory $z_M$ to get the probability distribution $\Pr(i|u, z_M)$ as defined in Equation (\ref{softmax_fun}), we only need to compute $L(i)-1$ times in the order of $\log |c(z_M)|$. Also hierarchical softmax does not increase the number of parameters to be estimated. 
Instead of estimating parameters of  $|c(z_M)|$ apps, we only need to estimate the parameters for $|c(z_M)|-1$ internal nodes.

\vspace{5pt}
\noindent{\bf Comments:} The binary tree built for hierarchical softmax is meant for computation efficiency purpose, which is different form the category tree $\Gamma$ used for structural choice modeling.

\subsection{Parameter Learning} 
After building the hierarchical softmax binary tree for each most outside subcategory node $z_M$ in the category hierarchy, the unique structural path for user $u$ to choose app $i$ is extended as 
$$\mathrm{path}_{i} = z_0 \rightarrow z_1  \cdots \rightarrow z_M \rightarrow n(i,1) \cdots \rightarrow n(i,L(i)).$$
We rewrite the log likelihood $\ell (\Theta)$ and get the following objective function 
\begin{equation*}
\begin{split}
    \mathcal O   = &  \sum_{u=1}^U \sum_{i \in \mathcal{D}_u}  \sum_{l=1}^{L(i)-1} \ln \Pr({u,n(i,l) \rightarrow n(i,l+1) }) \\
    & + \sum_{u=1}^U \sum_{i \in \mathcal{D}_u}\sum_{m=1}^{M} \ln \Pr(z_m|u, z_{m-1}) 
   + \sum_{\substack{m=1 \\\forall z\in \Gamma}}^{M} \ln \Pr(\qq_{z_m}|\qq_{z_{m-1}}, \sigma^2 \II)\\
   = & \sum_{u=1}^U \sum_{i \in \mathcal{D}_u}  \sum_{l=1}^{L(i)-1} \ln  \sigma \Big(\mathbb{S}\big(n(i,l+1) = \mathrm{left}(n(i,l))\big)\cdot y_{u, n(i,l)} \Big )\\
   & + \sum_{u=1}^U \sum_{i \in \mathcal{D}_u} \sum_{m=1}^{M} \ln \frac{\exp(b_{z_m} + \pp_u^\top \qq_{z_m})}{\sum_{z'\in c(z_{m-1})}\exp(b_{z'} + \pp_u^\top \qq_{z'})} \\
  & + \sum_{\substack{m=1 \\\forall z\in \Gamma}}^{M} \ln \mathcal N(\qq_{z_m}|\qq_{z_{m-1}}, \sigma^2 \II)
\end{split}
\end{equation*}
Note that here we have an updated set of parameters to estimate, namely, $\Theta =\{\pp_u, \qq_z, \qq_{n(i,l)}, b_z, b_{n(i,l)}\}$. Instead of estimating $\pp_i ~\mathrm{and} ~b_i ~\mathrm{for}~ i\in \mathcal I$, we estimate that of internal nodes $n(i,l)$ in the hierarchical softmax binary trees. 

We use stochastic gradient ascent method to update the latent factor variables. Stochastic gradient ascent (descent) has been widely used for many machine learning tasks~\cite{Bottou2010SGD}. The main process involves randomly scanning training instances and iteratively updating parameters. In each iteration, we randomly sample a  user-app adoption instance $\langle u, i, \mathrm{path}_{i}\rangle$, and we maximize $\mathcal O  (\Theta)$ using the following update rule for $\Theta$:
\begin{equation}
    \begin{aligned}
        \Theta = \Theta + \epsilon \cdot \frac{\partial \mathcal O  (\Theta)}{\partial \Theta},
   \end{aligned}
\end{equation}
where $\epsilon $ is a learning rate.

Specifically, given a user-app adoption instance $\langle u, i, \mathrm{path}_{i}\rangle$, the gradient with respect to $\pp_u$ is
\begin{equation}\label{update_pu}
\begin{split}
\small
     \frac{\partial \mathcal O }{\partial \pp_u} 
     = & \sum_{l=1}^{L(i)-1} \Big( \mathbf{1}_{l+1} - \sigma \big( y_{u,n(i,l)} \big ) \Big)\cdot \qq_{n(i,l)} \\
    & + \sum_{m=1}^{M} \Bigg( \qq_{z_m} - \frac{\sum_{z'\in c(z_{m-1})}\exp(b_{z'} + \pp_u^\top \qq_{z'})\cdot \qq_{z'}}{\sum_{z'\in c(z_{m-1})}\exp(b_{z'} + \pp_u^\top \qq_{z'})} \Bigg)
\end{split}
\end{equation}
Here $\mathbf{1}_{l+1}$ is an indicator function defined as
\begin{equation}\label{HS_indicator}
\mathbf{1}_{l+1} :=
\begin{cases} 
1 &\text{if } n(i,l+1) = \mathrm{left}(n(i,l)), \\
0 &\text{otherwise.}
\end{cases}
\end{equation}

Before moving to the internal nodes, let us define another indicator function $\mathbf{1}_{z\in \mathrm{path}_i}$ which is defined as
\begin{equation}
\mathbf{1}_{z\in \mathrm{Path}_i} :=
\begin{cases} 
1 &\text{if $z$ is in $\mathrm{path}_i$}, \\
0 &\text{other siblings nodes.}
\end{cases}
\end{equation}
Then, for each internal node $z \in \mathrm{path}_{i}$ and its siblings, we have the gradient with respect to $\qq_{z}$ as
\begin{equation}\label{update_q_z}
\begin{split}
 \frac{\partial \mathcal O }{\partial \qq_{z}}  
 & =\sum_{l=1}^{L(z)} \frac{\partial \ln \Pr (z|u, \pi(z))}{ \partial \qq_z } +  \sum_{\substack{m=1 \\\forall z\in \Gamma}}^{M}\frac{\partial  \ln \Pr(\qq_{z_m}|\qq_{z_{m-1}}, \sigma^2 \II)}{\partial \qq_z}\\
& = \mathbf{1}_{z\in \mathrm{path}_i}\cdot \pp_{u} - \frac{\exp(b_{z} + \pp_u^\top \qq_{z})\cdot \pp_{u}}{\sum_{z'\in c(z_{m-1})}\exp(b_{z'} + \pp_u^\top \qq_{z'})}\\
 & ~~~~- \frac{\qq_{z}- \qq_{\pi(z)}}{\sigma^2} -\frac{\sum_{z'\in c(z)}(\qq_{z}- \qq_{z'})}{\sigma^2}.
\end{split}
\end{equation}
Moreover, we have the gradient with respect to bias $b_{z}$ as
\begin{equation}\label{update_b_z}
\begin{split}
\frac{\partial \mathcal O }{\partial b_{z}} 
& = \mathbf{1}_{z\in \mathrm{Path}_i} - \frac{\exp(b_{z} + \pp_u^\top \qq_{z})}{\sum_{z'\in c(z_{m-1})}\exp(b_{z'} + \pp_u^\top \qq_{z'})}\\
\end{split}
\end{equation}

Finally, for each node level $l=\{ 1, 2, ...,L(i)-1 \}$ in the hierarchical softmax binary tree, we have the gradient with respect to $\qq_{n(i,l)}$ and $b_{n(i,l)}$ as 
\begin{align}
    \frac{\partial \mathcal O }{\partial \qq_{n(i,l)}} & =  \Big( \mathbf{1}_{l+1} - \sigma \big( y_{u,n(i,l)} \big ) \Big)\cdot \pp_u  \label{update_q_n_l}\\
\frac{\partial \mathcal O }{\partial b_{n (i,l)}} & =  \mathbf{1}_{l+1} - \sigma \big( y_{u,n(i,l)} \big ), \label{update_b_n_l}
\end{align}
where $\mathbf{1}_{l+1}$ is the  indicator function defined in Equation (\ref{HS_indicator}).
With gradients with respect to $\Theta =\{\pp_u, \qq_z, \qq_{n(i,l)}, b_z, b_{n(i,l)}\}$ being derived, we update $\Theta$  using stochastic gradient ascent rule $\Theta = \Theta + \epsilon \cdot \frac{\partial \mathcal O  (\Theta)}{\partial \Theta}$.
We summarize the parameter estimation procedure in Algorithm \ref{alg:Structural_choice}.

\begin{algorithm}
\DontPrintSemicolon 
\KwIn{category tree $\Gamma$, user app adoption observations $\mathcal D = \{(u, i)\}$, learning rate $\epsilon $.} 
\KwOut{optimal $\Theta =\{\pp_u, \qq_z, \qq_{n(i,l)}, b_z, b_{n(i,l)}\}$} 
\Begin{
\For{each most outside subcategory node $z_M$}{build a binary tree for hierarchical softmax}
Initialize $\Theta$ \;
\Repeat{convergence or reach max\_iter}{
    sample a user app adoption instance $\langle u, i, \mathrm{path}_{i}\rangle$ \;
    \tcp{update user latent factor}
    $\pp_u \leftarrow \pp_u + \epsilon \cdot \frac{\partial \mathcal O }{\partial \pp_u}$ (Equation (\ref{update_pu})) \;
    \tcp{update internal node latent factor}
    \For{each internal node $z \in \mathrm{path}_{i}$ and its siblings}{
    $\qq_{z} \leftarrow \qq_{z} + \epsilon \cdot \frac{\partial \mathcal O  (\Theta)}{\partial \qq_{z}}$ (Equation (\ref{update_q_z}))\;
    $b_{z} \leftarrow b_{z} + \epsilon \cdot \frac{\partial \mathcal O  (\Theta)}{\partial b_{z}}$ (Equation (\ref{update_b_z}))\;
    }
    \tcp{update hierarchical softmax binary tree node latent factor}
    \For{for each node level $l=\{1,...,L(i)-1\}$}{
    $\qq_{n(i,l)} \leftarrow \qq_{n (i,l)} + \epsilon \cdot \frac{\partial \mathcal O  (\Theta)}{\partial \qq_{n (i,l)}}$ (Equation (\ref{update_q_n_l}))\;
    $b_{n (i,l)} \leftarrow b_{n (i,l)} + \epsilon \cdot \frac{\partial \mathcal O  (\Theta)}{\partial b_{n (i,l)}}$ (Equation (\ref{update_b_n_l}))\;
    }
}
\Return $\hat{\Theta}$\;
} 
\caption{Structural User Choice Model Estimation} 
\label{alg:Structural_choice}
\end{algorithm}

\subsection{Complexity Analysis}
Note that in each iteration our structural user choice model has a linear time complexity $O\bigg( \left(\sum_{m=1}^M |L_m| + \log |c(z_M)|\right)\times |\mathcal D|\bigg)$, where $|\mathcal D|$ is the number of user-app adoption observations in the training dataset, $|L_m|$ is the number of categories  or subcategories in the category hierarchy level $m$, and  $\log |c(z_M)|$ is the logarithm of the number of apps under the most outside subcategory $z_M$, whose children nodes are apps. Therefore, the structural user choice model has the same complexity as the widely used latent factor models, which are usually linear to the number of observations $|\mathcal D|$. In most applications, value of $\left(\sum_{m=1}^M |L_m| + \log |c(z_M)|\right)$ will not be a large number. For example, in our app recommendation application with a dataset collected from Google Play, the worst case of $\left(\sum_{m=1}^M |L_m| + \log |c(z_M)|\right)$ is around $70$.

\section{Experiments}
\label{sec:exp}

This section presents an empirical evaluation of the performances of our model and previous methods. All the experiments are performed on a large-scale real-world app adoption dataset that we collected from Google Play.

\subsection{Dataset Collection}
The Google Play is a centralized marketplace where all apps are organized in a predefined category tree. Apps are organized into $27$ categories, and the category {\it Games} is further divided into $18$ subcategories. Also Google Play has both free and paid apps.  
Users can review (\emph{i.e.}, rate or like) apps on Google Play. A user's review about apps he/she used are publicly available. Once we obtain the Google ID of a user, we can locate all apps the user has reviewed. Therefore, we first obtained a list of Google user IDs from the data set shared from Gong et al.~\shortcite{gong2012evolution} and wrote a crawler to collect the list of apps that had been reviewed by these users. For each retrieved app, we crawled its category and subcategory information from Google Play.

We treated a user having adopted an app if the review score, whose value is from one to five, is greater or equal to three. 
After excluding users who have adopted less than $40$ apps to avoid cold start problem, we obtained a dataset with $52,483$ users, $26,426$ apps, and $ 3,286,156$ review observations. 
The resulting user-app adoption matrix has a sparsity as high as $99.76\%$ and each user adopts $62.61$ apps on average, which is a very small fraction of all the apps. 
Table \ref{dataset} shows some basic statistics of our dataset.

\begin{figure}[t!]
        \centering
        \begin{subfigure}[b]{0.25\textwidth}
                \includegraphics[width=1.08\textwidth]{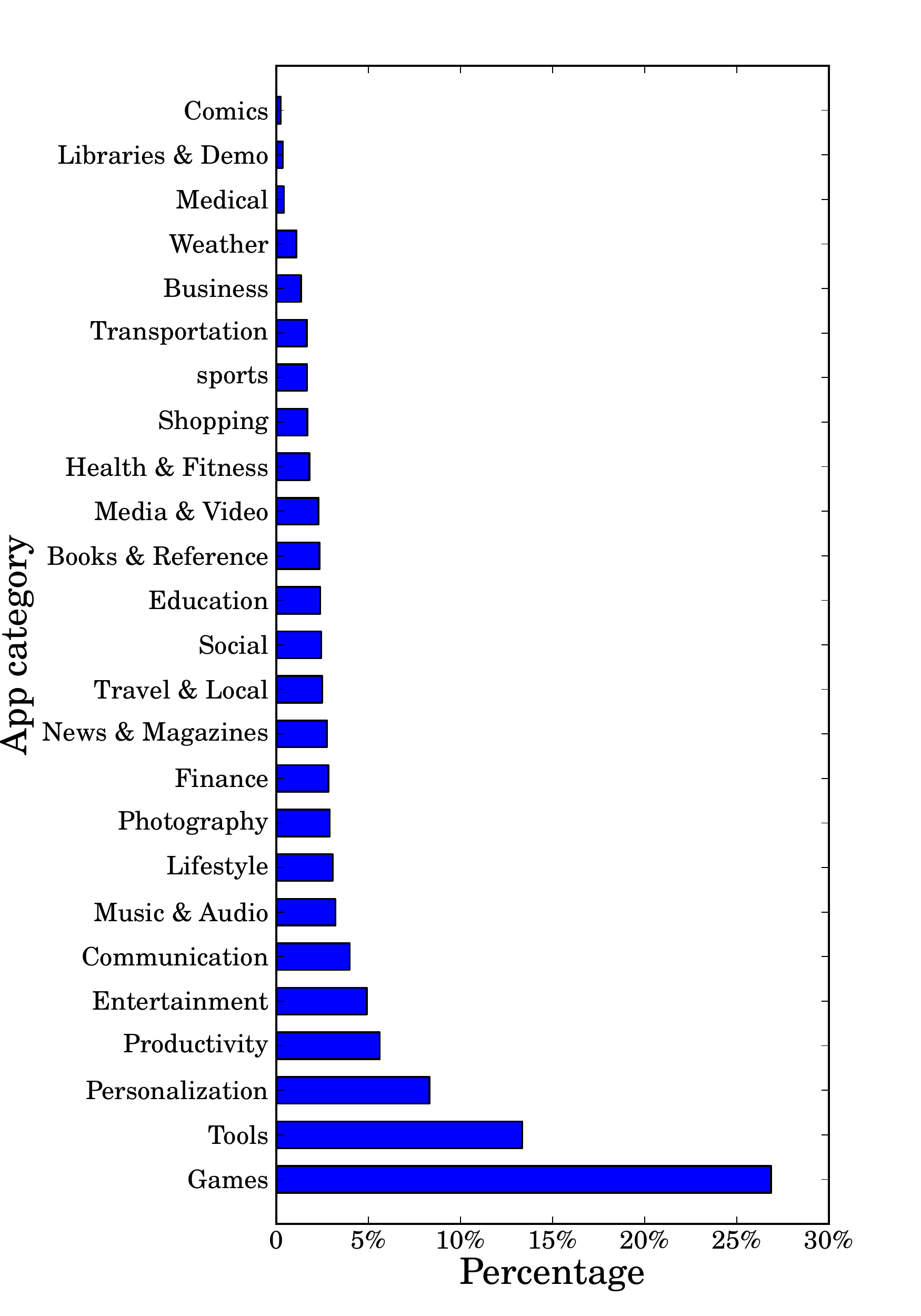}
                \caption{Category}
                \label{fig:cate_dist}
        \end{subfigure}%
        \begin{subfigure}[b]{0.25\textwidth}
                \includegraphics[width=1.08\textwidth]{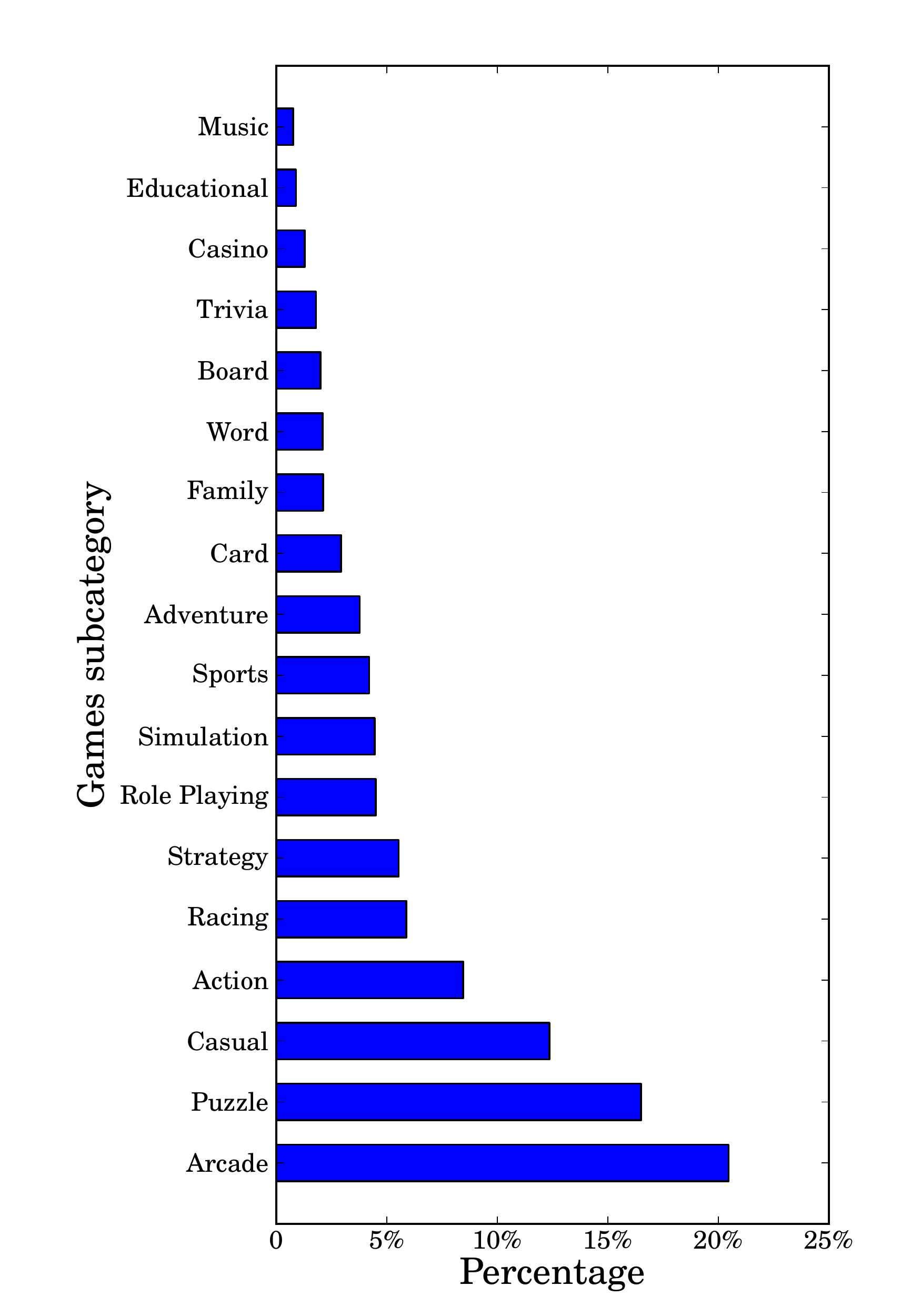}
                \caption{Games}
                \label{fig:games_dist}
        \end{subfigure}%
        \caption{App distributions in the category hierarchy. (a) App distributions in app categories, and (b) App distributions in {\it Games} subcategory.}\label{fig:tree_dist}
\end{figure}

Since only $11.11\%$ of all the apps in our dataset are paid apps, we do not distinguish between paid and free apps when constructing the hierarchical category  tree $\Gamma$. The $26,426$ apps are categorized into $25$ categories (the categories {\it Live Wallpaper} and {\it Widgets} defined by Google Play do not appear in our dataset). Figure \ref{fig:tree_dist} shows the detailed app distributions in different categories and the subcategories of Games. 
We observe that game apps take the highest percentage, accounting for $26.85\%$ of all the apps; {\it Arcade} ($20.46\%$), {\it Puzzle} ($16.50\%$), and {\it Casual} ($12.35\%$) are among the top three subcategories in {\it Games}, accounting for $49.31\%$ of all the game apps. 

\begin{table}[t]
\centering
\caption{Data Description}\label{dataset}
\vspace{-10pt}
\begin{tabular}{c c c c c c} \\ \thline
    \#users & \#apps &  \#observations & sparsity \\ \thline
    $52,483$ & $26,426$ & $3,286,156$ &  $99.76\%$  \\ \thline
      \end{tabular}
\end{table}

\subsection{Compared Approaches}\label{sec:baseline}
We compare our structural user choice model (SUCM) with the following recommendation models.
\begin{itemize}
\setlength{\itemsep}{5pt}
\item Logistic Latent Factor  Model (\textbf{LLFM})~\cite{Agarwal:KDD2009:RLFM}.  
LLFM was designed to model binary response using a cross-entropy loss function. In our problem, we adapt LLFM to solve the following optimization problem:
\begin{equation}
\begin{split}
 & \displaystyle \underset{\PP, \QQ, \mathbf b}{\operatorname{arg\,min}} \sum_{u, i\in \mathcal D}\ln \bigg(1+\exp\Big(-\left(\pp_u^\top \qq_i+ b_i\right)\Big)\bigg) \\ 
 & +  
\lambda_U \sum_{u\in \mathcal U} ||\pp_u||_2^2+\lambda_I \sum_{i \in \mathcal I} ||\qq_i||_2^2 + \lambda_b \sum_{i \in \mathcal I}b_i^2,
\end{split}
\end{equation}
where parameters $\lambda_U$, $\lambda_I$ and $\lambda_b$ are regularization weights for users, items, and item bias respectively.
\item Probabilistic Matrix Factorization with negative samples (\textbf{PMF}$_{\mathrm{Neg}}$)~\cite{PMF:NIPS08}. PMF is a standard latent factor model that is widely used for recommendations.  We adapt PMF to our problem, \emph{i.e.}, we solve the following  problem: 
\begin{equation}
\begin{split}
   & \displaystyle  \underset{\PP, \QQ, \mathbf b}{\operatorname{arg\,min}} \sum_{u, i\in \mathcal D}\bigg(y_{ui}-\left(\pp_u^\top \qq_i+ b_i\right)\bigg)^2 \\
   & + \lambda_U \sum_{u\in \mathcal U} ||\pp_u||^2+\lambda_I \sum_{i \in \mathcal I} ||\qq_i||^2+ \lambda_b \sum_{i \in \mathcal I}b_i^2.
\end{split}
\end{equation}
    However, we only have positive adopted instance $(u,i)$ which is treated as $y_{ui}=1$. Thus,  for each instance $(u,i)$, we sample a certain number  of negative instances $\{(u,j)\}$ and treat them as $y_{uj}=0$. We denote this modified PMF as \textbf{PMF}$_{\mathrm{Neg}}$. Note that \textbf{PMF}$_{\mathrm{Neg}}$ is similar to the sample based one-class collaborative filtering methods~\cite{Pan:one-class:2008,Hu2008:implicit_cf}.
\item \textbf{SVDFeature}~\cite{SVDfeature:2012}. SVDFeature is a feature-based latent factor model for recommendation settings with auxiliary information. We use category information as the auxiliary information for SVDFeature.
\item \textbf{LibFM}~\cite{libFM:2012,rendle2010factorization}.  LibFM is a software implementation for factorization machines (FM)~\cite{rendle2010factorization} that models all
    interactions between variables (\eg user, item and auxiliary information). 
    We also use category information as the auxiliary information and choose the 2-way FM.  
    One major difference between FM and SVDFeature is that SVDFeature only considers the interactions between user features and item features, while FM models all the interactions among all the available information.
\item Bayesian Personalized Ranking (\textbf{BPR})~\cite{Rendle:BPR:UAI2009}. BPR was first proposed to model personalized ranking with implicit feedback by treating observed user-item pairs as positive instances and sampling some  of the unseen user-item pairs as negative instances. Given the  preference triples $\DD = \{(u, i, j ) | i \succ j \}$, where $i \succ j$ indicates user $u$ prefers item $i$ than item $i$,  BPR aims at maximizing the following optimization criterion,
\begin{equation*}\label{BPR_obj}
\begin{split}
    & \displaystyle \underset{\PP, \QQ, \mathbf b}{\operatorname{arg\,max}}  ~ \left\{ \ln \prod_{(u, i, j)\in \DD} \Pr((u, i, j ) | i \succ_u  j )  \Pr(\Theta)\right\}  \\
    &= \displaystyle \underset{\PP, \QQ, \mathbf b}{\operatorname{arg\,max}} \left\{ \ln \prod_{(u, i, j)\in \DD}  \sigma(y_{ui} - y_{uj}|\Theta) \Pr(\Theta)\right\}
\end{split}
\end{equation*}
where $\sigma(\cdot)$ is the sigmoid function $\sigma(x) = \frac{1}{1 + e^{-x}}$, and $\Pr(\Theta)$ are Gaussian priors for the parameters.
\item Collaborative Competitive Filtering (\textbf{CCF})~\cite{Yang:sigir2011:CCF}: Given an offer set, CCF models  user-item choice behavior by encoding a local competition effect to improve recommendation performances. For each instance $(u,i)$, we sample a certain number of negative instances  to formulate the offer sets $\{(u, i, \mathcal A(i) )$ as described in~\cite{Yang:sigir2011:CCF}. Given collections of choice decision making records $\DD = \{(u, i, \mathcal A(i) )\}$, CCF estimates the latent factors and the item bias terms by solving following optimization problem
\begin{equation*}
\begin{split}
 & \displaystyle \underset{\PP, \QQ, \mathbf b}{\operatorname{arg\,min}} \Bigg\{ \sum_{(u, i, \mathcal A(i) )\in \mathcal D} \Bigg( \ln \Bigg[ \sum_{j \in \mathcal A(i)}  \exp \left(\pp_u^\top \qq_j+ b_j\right) \Bigg] \\
 &  - \left(\pp_u^\top \qq_i+ b_i\right) \Bigg) + \lambda_U \sum_{u\in \mathcal U} ||\pp_u||_2^2+\lambda_I \sum_{i \in \mathcal I} ||\qq_i||_2^2 \\ 
 & + \lambda_b \sum_{i \in \mathcal I}b_i^2 \Bigg\}.
\end{split}
\end{equation*}
\end{itemize}

\myparatight{Implementations, training, and testing} 
All  models are implemented with a stochastic gradient ascent/descent
optimization method with an annealing procedure to discount learning rate $\epsilon$ at the iteration nIter with $\epsilon^{\mathrm{nIter}} = \epsilon\frac{\nu}{\nu + \mathrm{nIter} - 1} $ by setting  $\nu = 50 $. The learning rate $\epsilon$ and the regularization weights are set by cross validation.  
All parameters are initialized by a Gaussian distribution $\mathcal N(0, 0.1)$. We randomly sample $80\%$ of adopted apps of each user as the training dataset, and we use the remaining adopted apps for testing.

\subsection{Evaluation Metrics}
In this implicit feedback app recommendation setting, we present each user with $N$ apps that have the highest predicted affinity values but are not adopted by the user in the training phase, 
and we evaluate different approaches based on which of these apps were actually adopted by the user in the test phase. More specifically, we adopt a variety of  widely used metrics to evaluate different approaches. In the following, we elaborate each metric.

\vspace{5pt}
\myparatight{\bf Precision and Recall}
Given a top-N recommendation list $C_{N,\mathrm{rec}}$, precision and recall are defined as
\begin{equation}
\begin{split}
\mathrm{Precision}@N =\frac{|C_{N,\mathrm{rec}}\bigcap C_{\mathrm{adopted}}|}{N}\\
\mathrm{Recall}@N =\frac{|C_{N,\mathrm{rec}}\bigcap C_{\mathrm{adopted}}|}{|C_{\mathrm{adopted}}|},
\end{split}
\end{equation}
where $C_{\mathrm{adopted}}$ are the apps that a user has adopted in the test data. The precision and recall for the entire recommender system are computed by averaging the precision and recall over all the users, respectively.

\vspace{5pt}
\myparatight {\bf F-measure}  F-measure balances between precision and recall. We consider the $F_\beta$ metric, which is defined as
\begin{equation}
F_\beta = (1 + \beta^2) \cdot \frac{\mathrm{Precision} \times \mathrm{Recall} }{ \beta^2 \cdot \mathrm{Precision} + \mathrm{Recall}}.
\end{equation}
where $\beta <1$ indicates more emphasis on precision than recall. In our experiments, we use $F_\beta$ metric with $\beta =0.5$. 

\vspace{5pt}
\myparatight {\bf Mean Average Precision}  Average precision (AP) is a ranked precision metric that gives larger credit to correctly recommended apps in higher positions. AP@N is defined as the average of precisions computed at all positions with an adopted app, namely, 
\begin{equation}
\mathrm{AP}@N =  \frac{\sum_{k=1}^N P(k) \times \mathrm{rel}(k)}{\min\{N, |C_{\mathrm{adopted}}|\} },
\end{equation}
where $P(k)$ is the precision at cut-off $k$ in the top-N list $C_{N,\mathrm{rec}}$, and $\mathrm{rel}(k)$ is an indicator function equaling $1$ if the app at rank $k$ is adopted, otherwise zero.
Finally, mean average precision (MAP@N)  is defined as the mean of the average precision scores for all users. 

\vspace{5pt}
\myparatight {\bf Normalized Discounted Cumulative Gain}  NDCG is a ranked precision metric that gives larger credit to correctly recommended apps in higher  positions. Specifically, the discounted cumulative gain (DCG) given a cut-off $N$ is calculated by
\begin{equation}
 \mathrm{DCG_{N}} = \sum_{i=1}^{N} \frac{ 2^{rel_{i}} - 1 }{ \log_{2}(i+1)},
\end{equation}
where $rel_{i}$ is  is the relevance score, which is binary. Then the NDCG@N is computed as $ \mathrm{NDCG@N} = \frac{\mathrm{DCG_{N}}}{\mathrm{IDCG_{N}}} $, where $\mathrm{IDCG_{N}}$ is the $\mathrm{DCG_{N}}$ value of the ideal ranking list.  The NDCG for the entire recommender system is computed by averaging the NDCG over all the users.

\begin{figure*}
        \centering
        \begin{subfigure}[b]{0.34\textwidth}
                \includegraphics[width=1.02\textwidth]{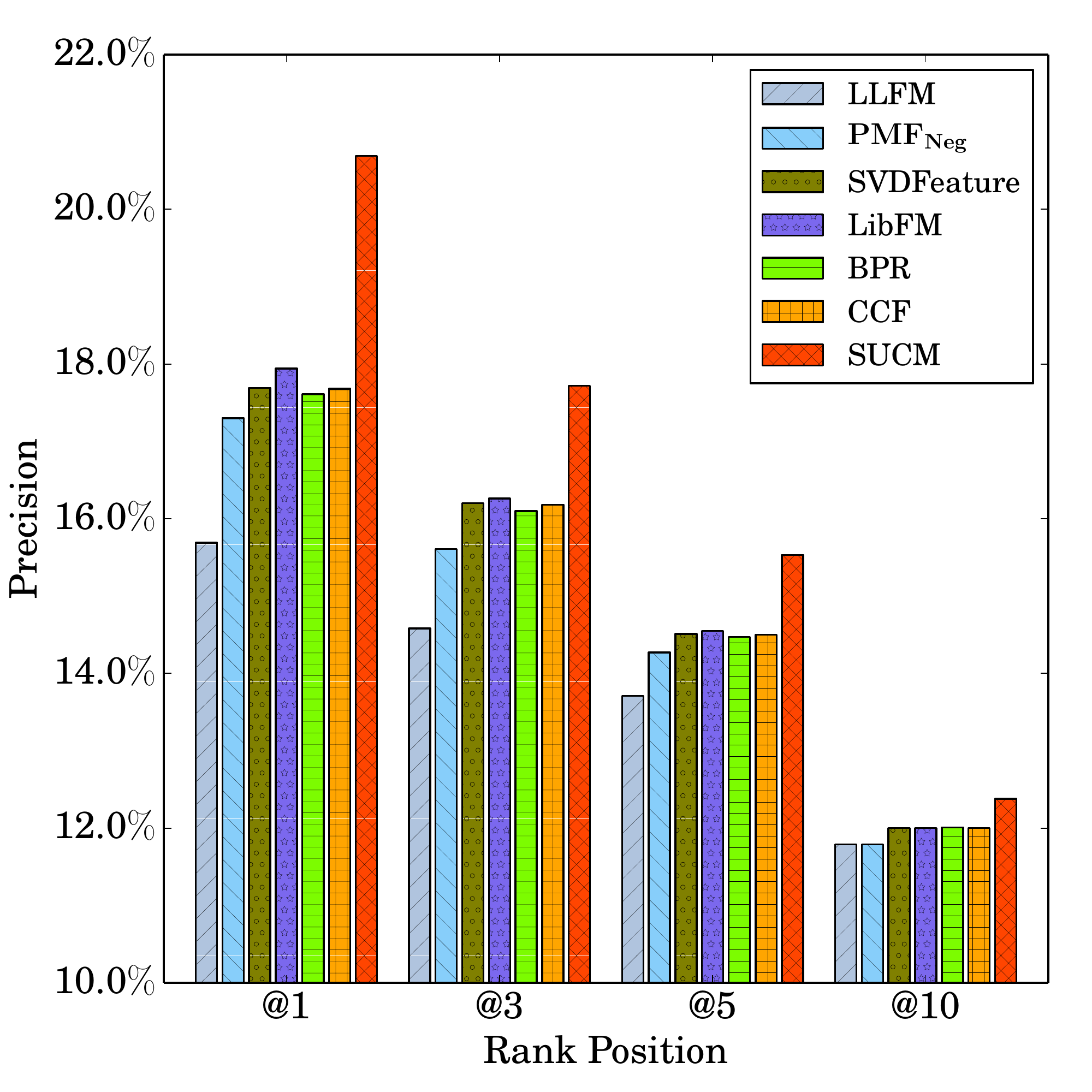}
                \caption{$K = 20$}
                \label{fig:bar_Pre_20}
        \end{subfigure}%
        \begin{subfigure}[b]{0.34\textwidth}
                \includegraphics[width=1.02\textwidth]{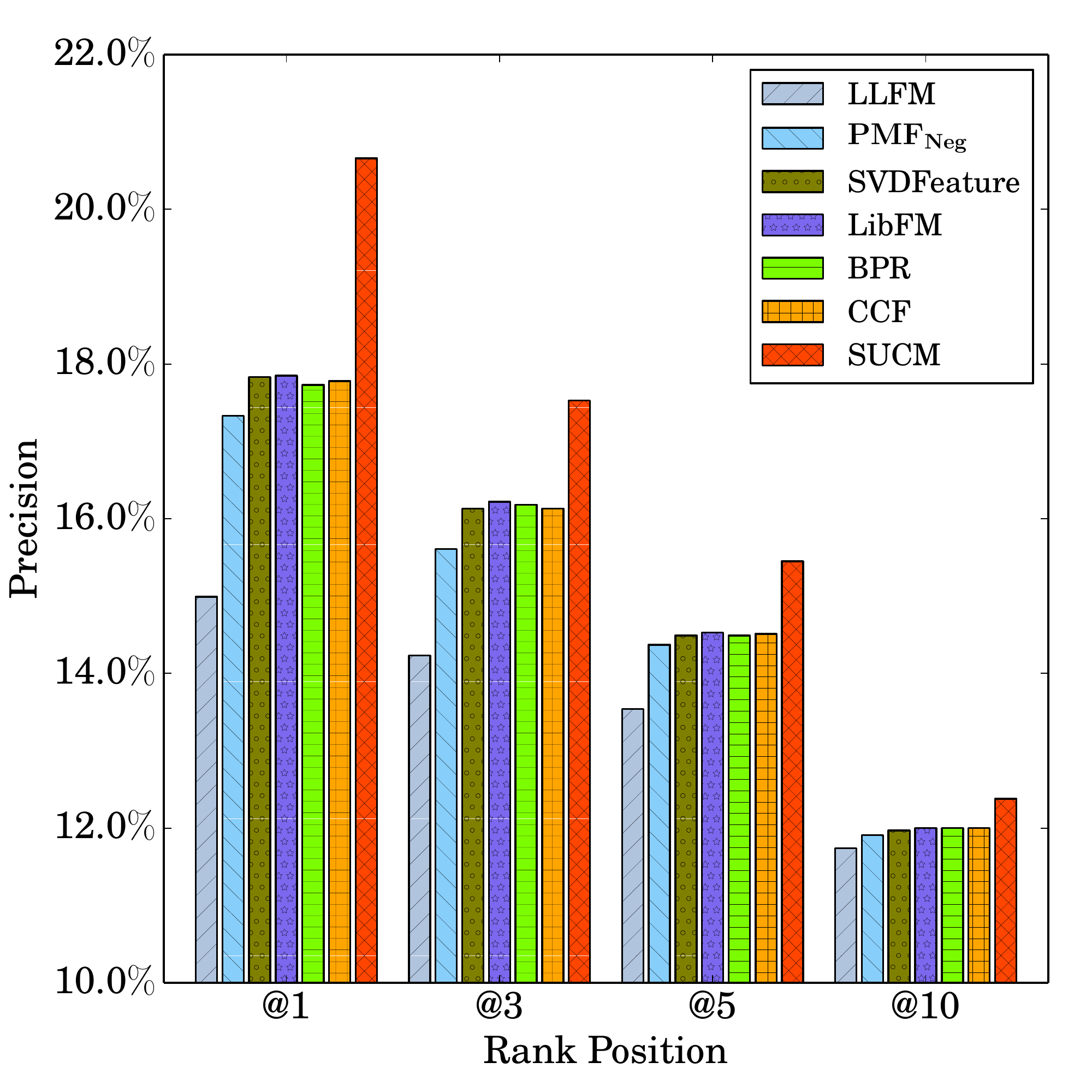}
                \caption{$K = 30$}
                \label{fig:bar_Pre_30}
        \end{subfigure}%
        \begin{subfigure}[b]{0.34\textwidth}
                \includegraphics[width=1.02\textwidth]{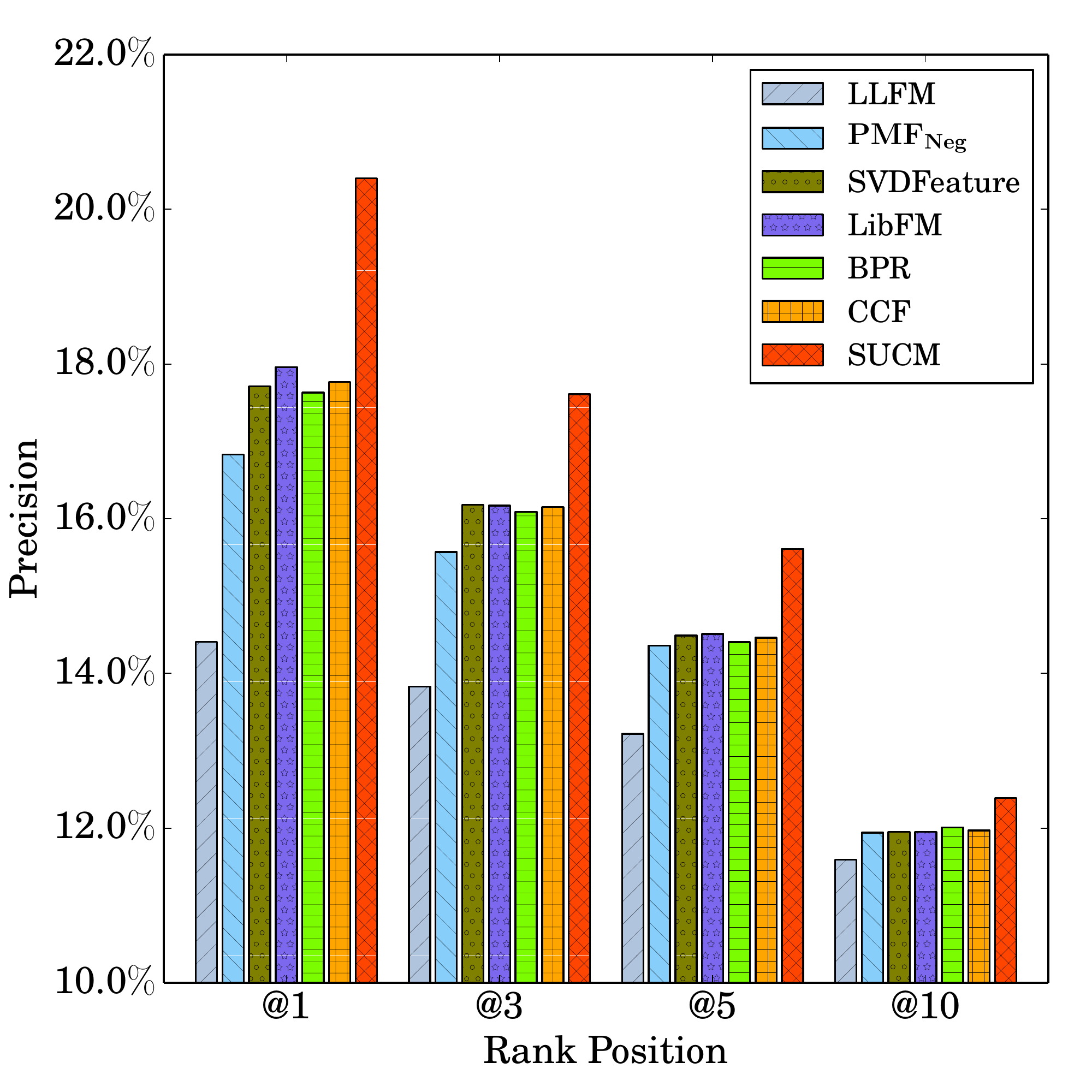}
                \caption{$K = 50$}
                \label{fig:bar_Pre_50}
        \end{subfigure}%
        \caption{Precision @N with  different latent dimensions $K$.}\label{fig:Precision}
\end{figure*}

\begin{figure*}
        \centering
        \begin{subfigure}[b]{0.34\textwidth}
                \includegraphics[width=1.02\textwidth]{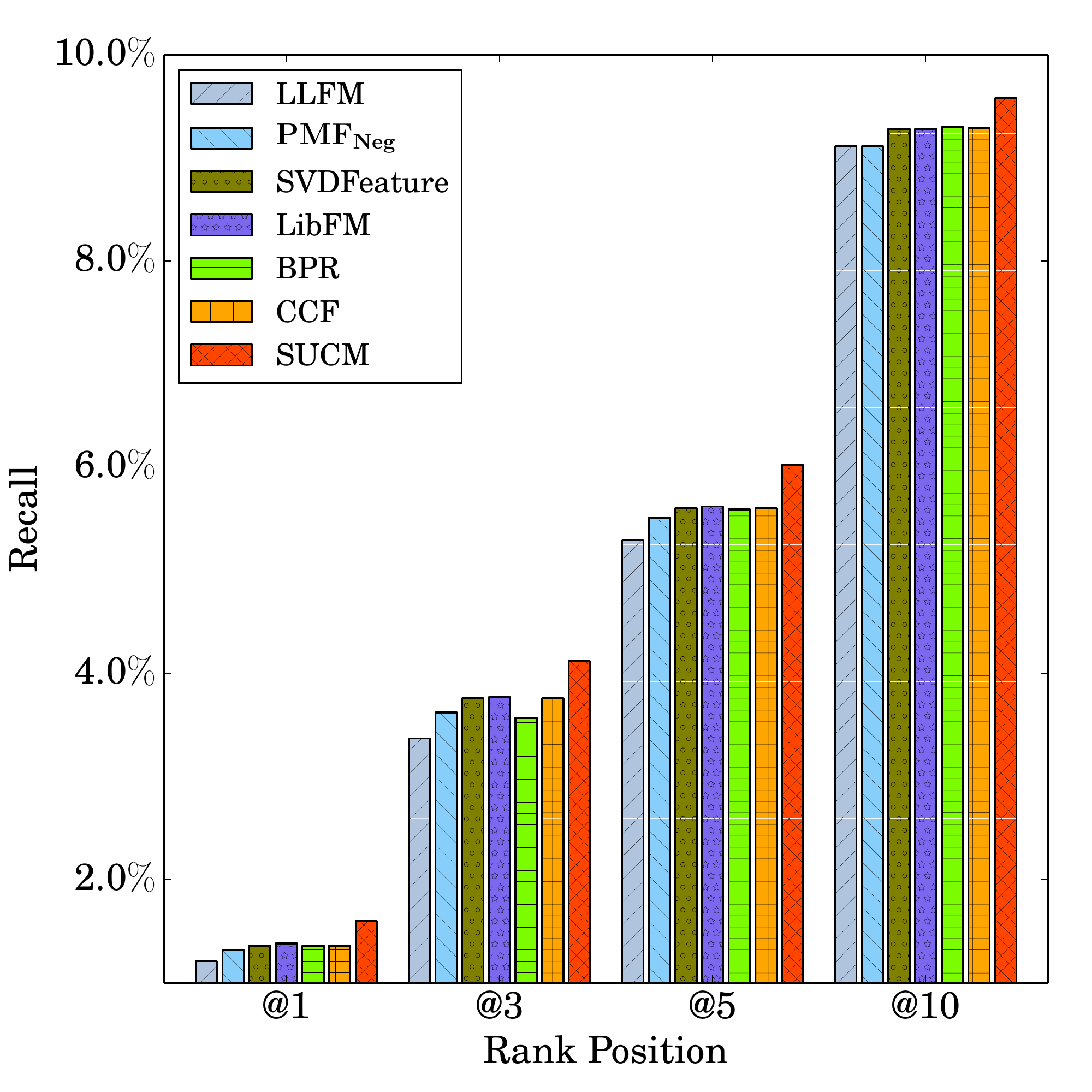}
                \caption{$K = 20$}
                \label{fig:bar_Rec_20}
        \end{subfigure}%
        \begin{subfigure}[b]{0.34\textwidth}
                \includegraphics[width=1.02\textwidth]{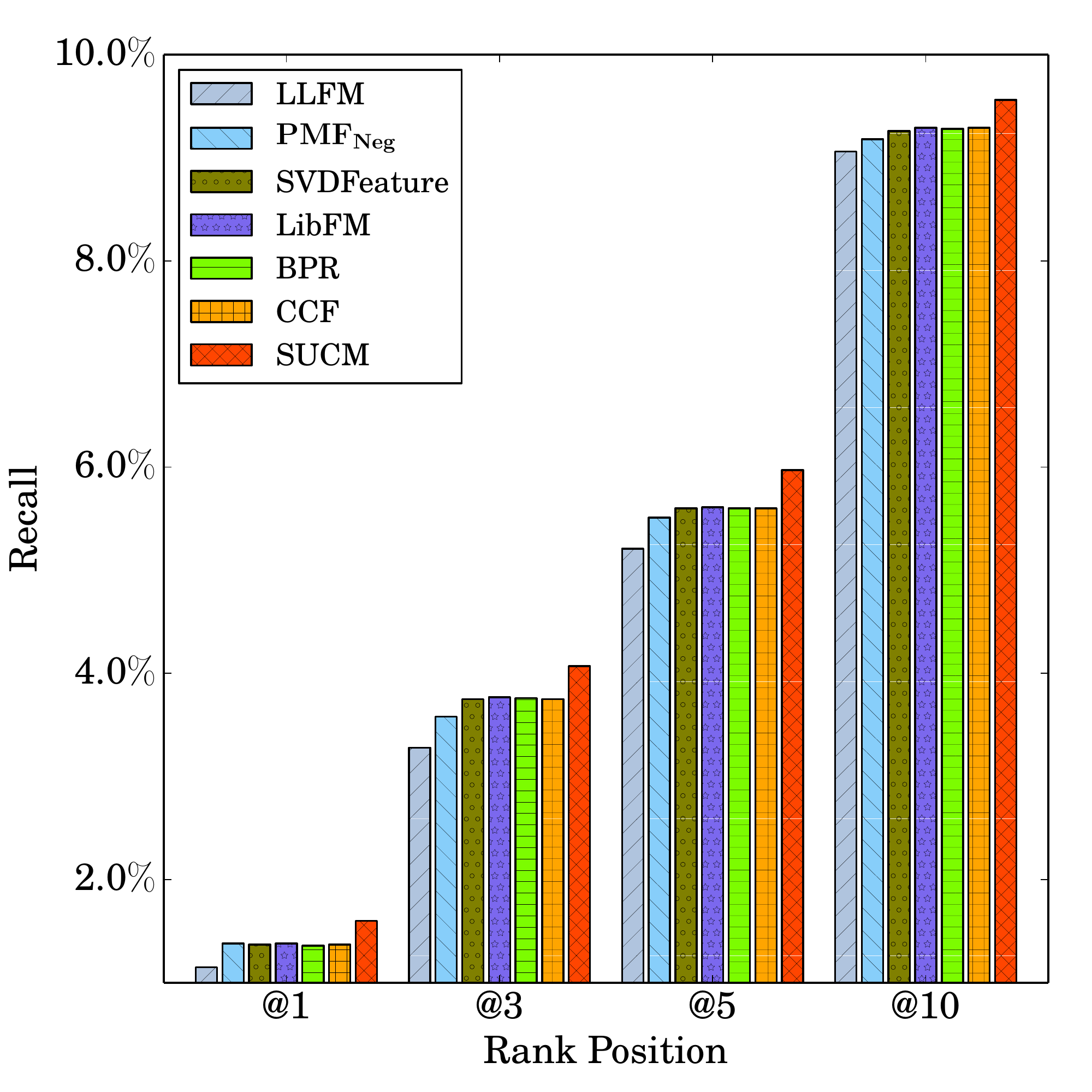}
                \caption{$K = 30$}
                \label{fig:bar_Rec_30}
        \end{subfigure}%
        \begin{subfigure}[b]{0.34\textwidth}
                \includegraphics[width=1.02\textwidth]{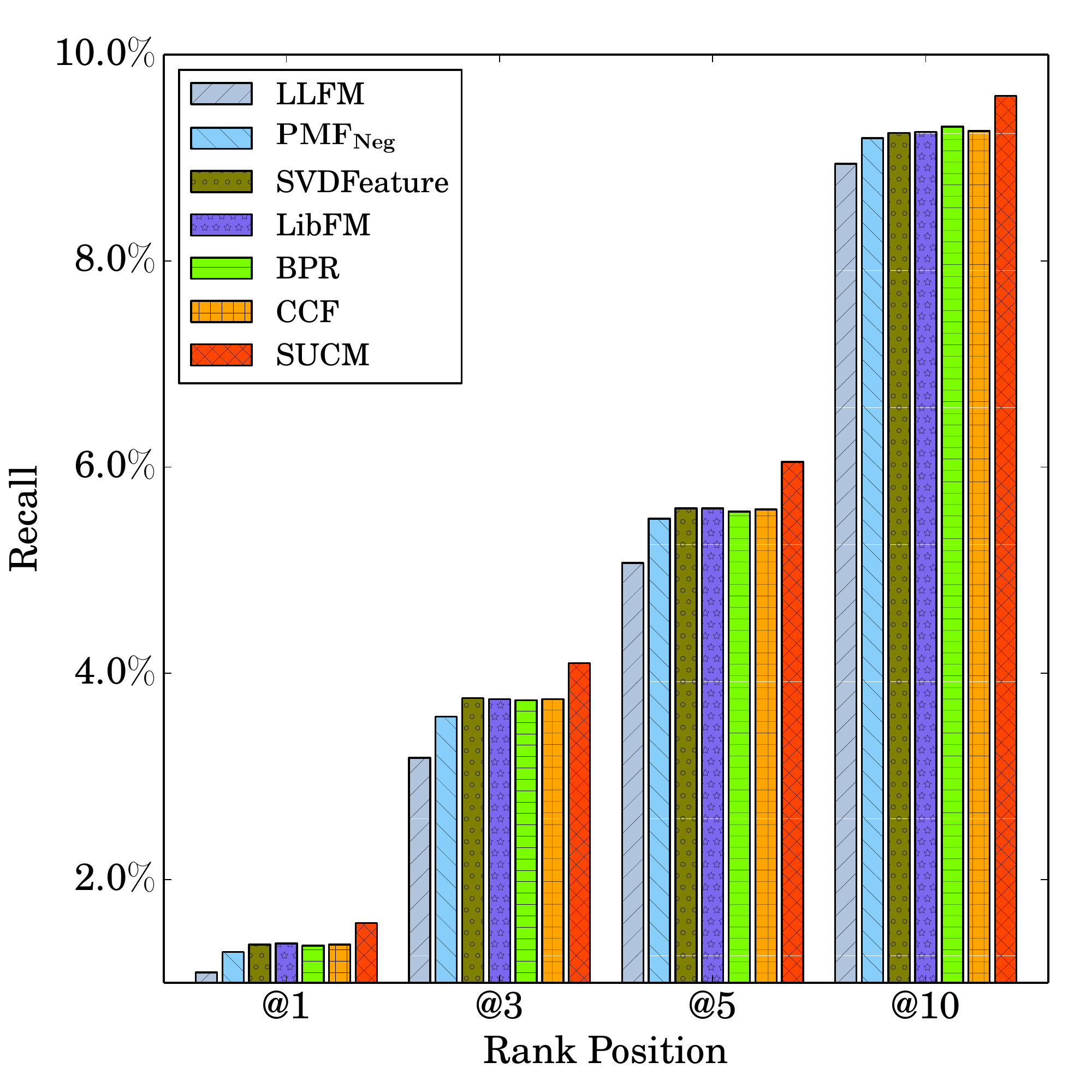}
                \caption{$K = 50$}
                \label{fig:bar_Rec_50}
        \end{subfigure}%
        \caption{Recall @N with  different latent dimensions $K$.}\label{fig:recall}
\vspace{-5pt}
\end{figure*}

\begin{figure*}
        \centering
        \begin{subfigure}[b]{0.34\textwidth}
                \includegraphics[width=1.02\textwidth]{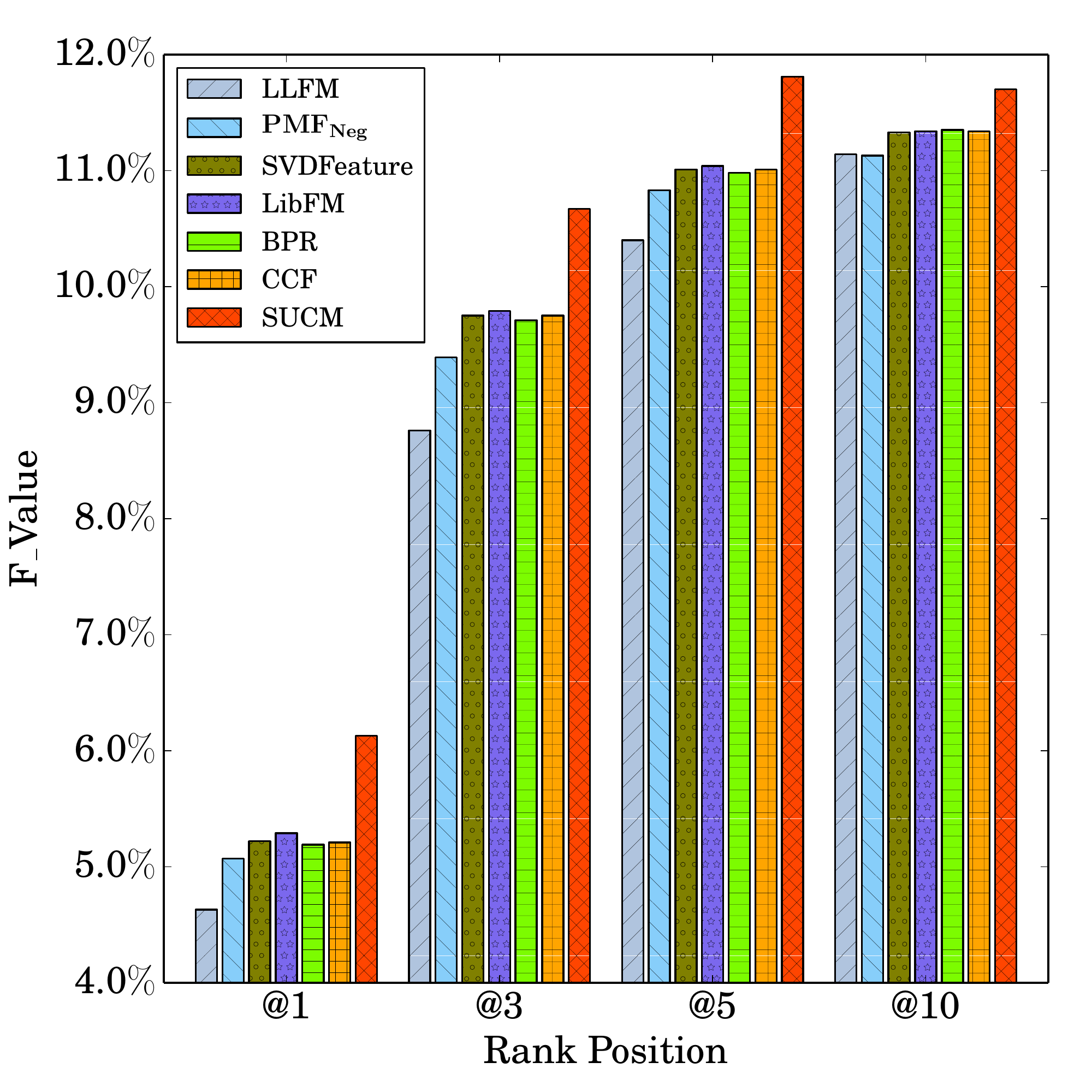}
                \caption{$K = 20$}
                \label{fig:bar_F_20}
        \end{subfigure}%
        \begin{subfigure}[b]{0.34\textwidth}
                \includegraphics[width=1.02\textwidth]{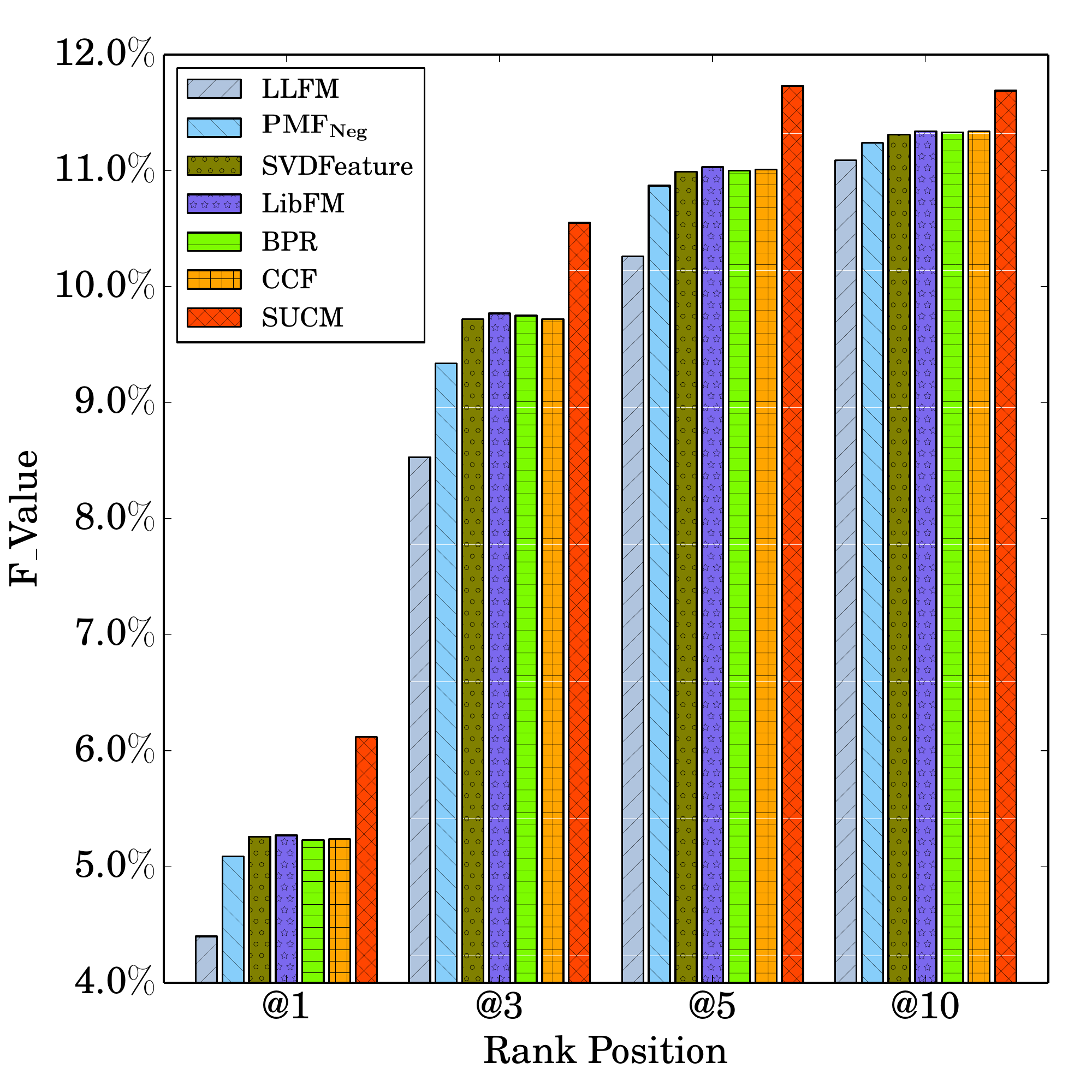}
                \caption{$K = 30$}
                \label{fig:bar_F_30}
        \end{subfigure}%
        \begin{subfigure}[b]{0.34\textwidth}
                \includegraphics[width=1.02\textwidth]{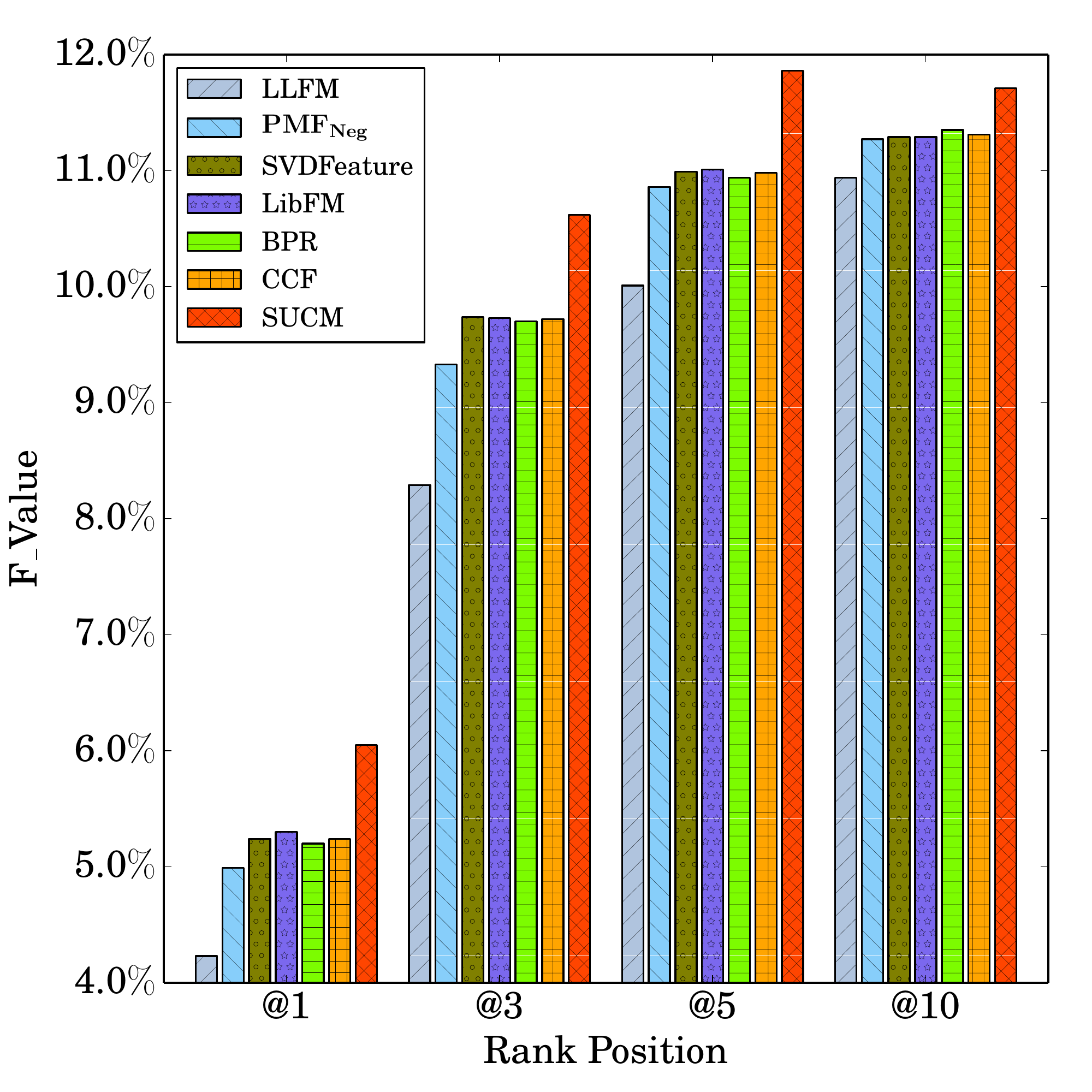}
                \caption{$K = 50$}
                \label{fig:bar_F_50}
        \end{subfigure}%
        \caption{F-measure $\mathrm{F}_{\beta}$@N with  different latent dimensions $K$ ($\beta = 0.5$).}\label{fig:F_measure}
\end{figure*}

\subsection{Performance Comparisons}\label{sec:comp}
In this subsection, we present the performance comparisons on top-N performances between our proposed SUCM and the baseline methods. 
We compare various approaches with three latent dimensions $K = 20$, $K = 30$, and $K = 50$, and four top-N values $N = 1, 3, 5, 10$.

Figure \ref{fig:Precision}, Figure \ref{fig:recall}, and  Figure \ref{fig:F_measure} respectively show the precision@N, recall@N, and $F_\beta$@N
of all compared approaches on our dataset. We find that our approach consistently and substantially outperforms the previous methods for different $N$ and different $K$. Moreover, we observe that negative sampling based methods PMF$_{\mathrm{Neg}}$ and BPR outperform LLFM which considers only positive instances for all the three considered number of latent dimensions. This is because LLFM polarizes towards the positive response values, and the learned recommendation model would predict positive for almost all unseen items and yield poor ranking performances. However, PMF$_{\mathrm{Neg}}$ and BPR mitigate the issue of LLFM via sampling unseen items as negative instances. Although PMF$_{\mathrm{Neg}}$ and BPR achieve close performances for top-1 recommendations, BPR works slightly better than PMF$_{\mathrm{Neg}}$ for top-3, top-5, and top-10 recommendations. Moreover,  CCF further slightly outperforms BPR in most cases. This is because CCF captures the local competition by a softmax of the chosen item over the offer set. Our proposed SUCM further improves upon CCF with significant margins for all the three evaluation metrics. For example, SUCM improves upon CCF with around $3\%$ in terms of top-1 precision. 

Besides, previous work ~\cite{SVDfeature:2012,libFM:2012,rendle2010factorization} showed that 
latent factor based recommendation could be improved by incorporating auxiliary information such as item features and context information. However, we observe that, by treating category information as auxiliary information, these methods (\eg SVDFeature and LibFM) can only gain marginal improvements in terms of top N recommendation performances. Compared with counterpart method PMF$_{\mathrm{Neg}}$ without auxiliary information, SVDFeature can only gain around 0.4\% improvement  and LibFM can only gain around 0.6\% improvement in terms of top-1 precision respectively. 
We argue that category information, treated as auxiliary feature, is not fine-grained enough to discriminate user preferences. Quite differently, SUCM  leverage the hierarchical structure of category information to better profile user interest preferences.

\vspace{5pt}
Precision, recall, and F-measure  do not consider the ranking positions of correctly recommended apps. So we further adopt MAP and NDCG  to provide more  fine-grained understanding of these recommendation approaches. Intuitively,   MAP and NDCG give larger credits to correctly recommended apps that are in higher  ranking positions. Table \ref{Tab:MAP} and Table \ref{Tab:NDCG} respectively show the MAP@N and NDCG@N of all compared approaches. Again, we observe consistent and substantial improvements of our  SUCM upon previous methods.

\begin{table*}[th]
\caption{MAP@N with  different latent dimensions $K$.}\label{Tab:MAP}
\addtolength{\tabcolsep}{-1pt}
\centering 
\begin{tabular}{c  c l l l l l l l}
    \thline 
K & MAP  & LLFM & PMF$_{\mathrm{Neg}}$  &   SVDFeature & LibFM & BPR & CCF & SUCM \\
\thline
\multirow{4}{*}{20}
   & @1     & 15.69\%   & 17.30\% & 17.69\% & 17.94\% & 17.61\%   & 17.68\%   & \bf 20.69\%   \\
   & @3     & 10.38\%   & 11.29\% & 11.72\% & 11.82\% & 11.64\%   & 11.69\%   & \bf 13.40\%   \\
   & @5     & 8.17\%   & 8.73\% & 9.00\% & 9.07\% & 8.95\%   & 8.99\%   & \bf 10.11\%   \\ 
   & @10    & 5.49\%   & 5.69\% & 5.85\% &  5.88\%  & 5.84\%   & 5.85\%   & \bf 6.40\%  \\ 
   \thline
\multirow{4}{*}{30}
   & @1     & 14.99\%   & 17.33\% & 17.83\% & 17.85\% & 17.73\%   & 17.78\%   & \bf 20.66\%   \\
   & @3     & 10.07\%   & 11.46\% & 11.71\% &  11.78\%  & 11.72\%   & 11.70\%   & \bf 13.23\% \\
   & @5     & 7.97\%   & 8.89\% & 9.01\% & 9.05\% & 9.00\%   & 9.00\%   & \bf 9.97\%   \\ 
   & @10    & 5.38\%   & 5.80\%  & 5.85\% &  5.87\% & 5.86\%   & 5.86\%   & \bf 6.33\%   \\ 
   \thline
\multirow{4}{*}{50}
   & @1     & 14.41\%   & 16.83\% & 17.71\% & 17.96\% & 17.63\%   & 17.77\%   & \bf 20.40\%   \\
   & @3     & 9.71\%   & 11.23\%  & 11.74\% & 11.77\% & 11.67\%   & 11.72\%   & \bf 13.25\%   \\
   & @5     & 7.71\%   & 8.77\% & 9.01\% & 9.03\% & 8.94\%   & 8.99\%   & \bf 10.05\%   \\ 
   & @10    & 5.25\%   & 5.76\% & 5.84\% & 5.86\% & 5.84\%   & 5.85\%   & \bf 6.36\%   \\ 
   \thline
\end{tabular}
\end{table*}

\begin{table*}[th]
\caption{NDCG@N with  different latent dimensions $K$.}\label{Tab:NDCG}
\addtolength{\tabcolsep}{-1pt}
\centering 
\begin{tabular}{c  c l l l l l l l}
\thline 
K & NDCG  & LLFM & PMF$_{\mathrm{Neg}}$  & SVDFeature & LibFM & BPR & CCF & SUCM \\
\thline
\multirow{4}{*}{20}
   & @1     & 15.69\%   & 17.30\%  & 17.69\% & 17.94\% & 17.61\%   & 17.68\%   & \bf 20.69\% \\ 
   & @3     & 14.83\%   & 15.96\%  & 16.54\% & 16.64\% & 16.45\%   & 16.52\%   & \bf 18.40\% \\ 
   & @5     & 14.17\%   & 14.94\%  & 15.29\% & 15.36\% & 15.23\%   & 15.28\%   & \bf 16.72\% \\ 
   & @10    & 12.69\%   & 12.99\%  & 13.27\% & 13.30\% & 13.26\%   & 13.26\%   & \bf 14.11\% \\ 
   \thline
\multirow{4}{*}{30}
   & @1     & 14.99\%   & 17.33\%  & 17.83\% & 17.85\% & 17.73\%   & 17.78\%   & \bf 20.66\% \\ 
   & @3     & 14.41\%   & 16.00\%  & 16.52\% & 16.59\% & 16.54\%   & 16.52\%   & \bf 18.24\% \\ 
   & @5     & 13.89\%   & 15.04\%  & 15.29\% & 15.34\% & 15.28\%   & 15.30\%   & \bf 16.61\% \\ 
   & @10    & 12.53\%   & 13.10\%  & 13.26\% & 13.29\% & 13.27\%   & 13.28\%   & \bf 14.07\% \\ 
   \thline
\multirow{4}{*}{50}
   & @1     & 14.41\%   & 16.83\%  & 17.71\% & 17.96\% & 17.63\%   & 17.77\%   & \bf 20.40\% \\ 
   & @3     & 13.96\%   & 15.87\%  & 16.54\% & 16.58\% & 16.47\%   & 16.52\%   & \bf 18.23\% \\ 
   & @5     & 13.51\%   & 14.96\%  & 15.28\% & 15.33\% & 15.21\%   & 15.26\%   & \bf 16.68\% \\ 
   & @10    & 12.29\%   & 13.08\%  & 13.24\% & 13.27\% & 13.26\%   & 13.25\%   & \bf 14.07\% \\ 
   \thline
\end{tabular}
\end{table*}

\vspace{5pt}
\myparatight{Summary} Through extensive evaluations, we found that our method SUCM consistently and substantially outperforms previous methods in terms of a variety of evaluation metrics. We argue that SUCM achieves this performance gain by learning fine-grained user preferences via leveraging the hierarchical category tree of apps and capturing the competitions between apps.

\section{Related Work}
Our work is related with two research fields: personalized recommendation methodology and mobile app recommendation. 

\myparatight{Recommendation methodology} 
The most popular model-based approaches are based on the latent factor models \cite{PMF:NIPS08,Koren:SVDRec2009,Agarwal:KDD2009:RLFM,wu2016cdae}.
For the binary implicit feedback setting, models such as LLFM use cross-entropy loss~\cite{Agarwal:KDD2009:RLFM}, but it is still apt to obtain an estimator that would polarize toward the positive response values, thus leading to limited top N performances.
Negative sampling provides an alternative by sampling a certain number of unseen items as negative samples. Then standard latent factor models such as probabilistic matrix factorization (PMF)~\cite{PMF:NIPS08} can be adopted. Hu \etal~\shortcite{Hu2008:implicit_cf} proposed to treat implicit data as indication of positive and negative preference associated with vastly varying confidence levels on the objective function. Pan \etal~\shortcite{Pan:one-class:2008} used a similar strategy by applying weighted low rank approximation. 
Instead of optimizing point-wise loss function, Bayesian Personalized Ranking (BPR) \cite{Rendle:BPR:UAI2009} optimizes a pairwise loss function to preserve the relative order of items for each user.

There are few works that adopt discrete choice models to model user item choices for recommendation~\cite{Yang:sigir2011:CCF} and for geographic ranking~\cite{Kumar:wsdm2015:GeoCHoice}. 
Discrete choice models~\cite{Luce:ChoiceAxiom1959,McFadden:choide:1973} are built on established theories on consumer preferences and utility, and have been widely used for understanding consumer behavior in different application domains such as travel~\cite{Ben1985discrete}, transportation~\cite{Train1978validation}, and brand choice~\cite{guadagni1983logit}. 
Based on discrete choice model, Yang \etal~\shortcite{Yang:sigir2011:CCF} proposed a collaborative competitive filtering (CCF) model to learn user-item choice to improve recommendation performance.  Given users' interaction with offer sets, CCF models user choice by a softmax function of the chosen item over the offer set. In this sense, CCF can also be categorized into sampling based method with samples given in the offer set. Though trying to model user choice process, CCF does not consider the structural dependence between items to be chosen for users, thus the choice model in CCF is ``flat'' rather then structural. We extend the ``flat'' choice model into structural user choice model to capture fine grained user preferences for mobile app recommendation.


\vspace{5pt}
Recently, there are a few works to explore the item hierarchy and side information for recommendation. For instance, Kanagal \etal~\shortcite{Kanagal:Taxonomy:vldb2012} and Ahmed \etal~\shortcite{Ahmed:wsdm2013:latent_hierarch} proposed to learn user preferences with additional hierarchical item relationship and other side information such as brand and temporal purchase  sequence.  Instead of using the predefined item hierarchy, some other
works also try to learn the item taxonomy~\cite{zhang2014taxonomy}. However, these works did not consider the competitions among similar items or among similar categories/subcategories.  We do not compare our methods with them because they utilized more side information such as item brands and item semantics, but it is an interesting future work for us to extend our framework to incorporate similar side information in the domain of app recommendation. Besides, Ziegler \etal~\shortcite{Ziegler:2005:DiveRec} utilized  taxonomy information to balance and diversify personalized recommendation lists. However, our work different from Ziegler \etal~\shortcite{Ziegler:2005:DiveRec} in both the purpose and the way of utilizing taxonomy information.


\myparatight{Mobile app recommendation} Recently, app recommendation has drawn an increasing number of attentions. Different from 
other domains such as movies~\cite{Yehuda:Netflix:2007}, musics~\cite{aizenberg2012build}, and point-of-interests~\cite{BinKDD13,Bin:TKDE:2015}, app recommendation has its own characteristics. Yin \etal~\shortcite{Yin:AppRec:WSDM2013} considered a trade-off between satisfaction and temptation for app recommendation with a special focus on the case that a user would like to replace an old app with a new one. 
Similarly, \cite{Lin:sigir:2014,Lin:phdthesis} considered app versions to improve app recommendation by  incorporating features distilled from version descriptions.  
Karatzoglou \etal \shortcite{Karatzoglou:contextAppRecCIKM12} provided a context-aware recommendation using tensor factorization by including context information such as location, moving status, and time.  Woerndl \etal \shortcite{woerndl2007hybrid} applied a hybrid method  for context-aware app recommendation. 
To address the cold-start problem for app recommendation, \cite{Lin:appRec:sigir13,Lin:phdthesis} proposed to leverage  side information from Twitter. Specifically, information of followers of an app's official Twitter account is collected and utilized to model the app, providing an estimation about which users may like the app. 
Davidsson \etal \shortcite{Davidsson:2011:UIF}  presented a context-based recommender prototype  for cold-start user users. 
Zhu {\it et al.} \shortcite{Zhu:AppRank:kdd14} proposed a mobile app ranking system by considering both the app's popularity and security risks.  More recently, Liu \etal~\shortcite{Bin:WSDM15} studied personalized app recommendation by reconciling user functionality preferences and user privacy preferences. 
Baeza-Yates \etal \shortcite{Baeza-Yates:WSDM2015} proposed a method to predict  which app a user is going to use by leveraging spatio-temporal context features, and
Park \etal \shortcite{Park:SigIR2015} proposed a method to improve the accuracy of mobile app retrieval by jointly modeling app descriptions and user reviews using topic model. 
However, these works are orthogonal to ours because they use other auxiliary information such as app versions, app satisfaction and temptation, and app privacy, while our work focuses on leveraging app taxonomy to model structural user choices among competing apps.


\section{Conclusion and Future Work}

In this paper,  we proposed a novel structural user choice model to learn fine-grained user preferences via leveraging the tree hierarchy of apps and capturing competitions between apps for app recommendation. Specifically, given all apps in a centralized mobile app market  organized as a category tree, we represented the structural user choice as a unique choice path, starting from the root till the leaf node where user makes an app adoption decision, over the category hierarchy.  Then we captured the structural choice procedure by cascading user preferences over the choice path through a novel probabilistic model. We also designed an efficient learning algorithm to estimate the model parameters. Moreover, we collected a real-world large-scale user-app adoption dataset from Google Play and used it to evaluate our method and various previous methods. Our results demonstrated that our method achieved consistent and substantial performance improvements over previous methods. 

There are a few interesting future directions that are worth exploring. 
1) Human-induced taxonomies are usually noisy and incomplete, and they do no evolve with a change in user demographics or product inventory. Zhang \etal \cite{zhang2014taxonomy} proposed a probabilistic model that is able to automatically discover the taxonomies from online shopping data. An interesting future work is to explore a unified model that could jointly learn the taxonomy and the structural choice model.
2) There are plenty of user reviews in the Google Play store. Several previous works \cite{zhang2014explicit,wu2015flame,chen2016sigir} have shown that user reviews provide more detailed information on why a user gives an item a specific rating. Integrating our model with user reviews helps us better understand fine-grained user preferences. 
3) We are also interested in adopting our model to other domains where items are also hierarchically categorized.

\balance
\bibliographystyle{abbrv}
{
\small
\bibliography{refs}
}

\end{document}